\documentclass[twocolumn,showpacs,showkeywords,preprintnumbers,amsmath,amssymb]{revtex4}

\usepackage{graphicx}
\usepackage{dcolumn}
\usepackage{bm}
\usepackage{pst-all,mathrsfs}
\psset{unit=1cm,fillcolor=white,linewidth=.3mm,arrowsize=5pt,coilarm=.25cm,coilwidth=.25cm}
\newrgbcolor{brown}{0.50 0.25 0.00}
\DeclareRobustCommand\openone{\leavevmode\hbox{\small1\normalsize\kern-.33em1}}%
\newcommand{\ket}[1]{\ensuremath{|#1\rangle}}%
\newcommand{\tr}[2][\mbox{}]{\textrm{tr}_{#1}\left[#2\right]}%
\newcommand{\vmedio}[1]{\ensuremath{\langle#1\rangle}}%
\newcommand{\definido}{\stackrel{\scriptscriptstyle\textrm{def}}{\displaystyle =}}%
\newcommand{\bra}[1]{\ensuremath{\langle#1|}}
\newcommand{\braket}[2]{\langle#1|#2\rangle}
\newcommand{\vi}{\hat{\imath}}
\newcommand{\vj}{\hat{\jmath}}
\newcommand{\vk}{\hat{k}}
\renewcommand{\Re}{\textnormal{Re}}

\DeclareRobustCommand{\exponencial}{\leavevmode\hbox{\LARGE\textnormal{e}\normalsize}}
\renewcommand{\exp}{\exponencial}

\begin{document}
\preprint{APS/123-QED}
\title{Exact analytical solution of Entanglement of Formation and Quantum Discord for Werner state and
Generalized Werner-Like states}%
\author{S. D\'{i}az-Sol\'{o}rzano}%
\email{sttiwuerdiaz@usb.es}
\author{E. Castro}
\altaffiliation[Electronic address:\space]{\texttt{ecastro@usb.ve}}%
\affiliation{%
Physics Department, Quantum Information and Communication Group, Sim\'{o}n Bol\'{i}var
University, Apdo. 89000, Caracas 1086, Venezuela.
}%

\date{\today}

\begin{abstract}\noindent%
We obtained analytical expressions for Entangled of Formation (EoF) and Quantum Discord
(QD) of Werner states and Generalized Werner-Like states. The optimization problem
involved under the exact analytical form is obtained for both quantities. In order to
illustrate the importance of our results we studied the EoF and the QD of these states.
Using discrete formulation of continue states with the $f-$deformed coherent states
obtained as deformed annihilation operator coherent states and as deformed displacement
operator coherent states. The EoF and QD of bipartite Werner-Like states $f-$deformed
coherent states are studied for the P\"{o}schl-Teller, Morse and quantum dot deformed
potentials. The result obtained are compared with the case of bipartite Werner-Like
coherent states.
\end{abstract}


\maketitle%
\section{Introduction}\label{sec:intro}%
\vspace{-.5cm} Quantum correlations lie in the foundation of quantum mechanics and are
the heart of quantum information science. They are important to study the differences
between the classical and quantum worlds because, in general, the quantum systems can be
correlated in ways inaccessible to classical objects. The research on quantum correlation
measures were initially developed on the entanglement-separability paradigm~\cite{Amico}
(and the references therein). However, it is well known that entanglement does not
account for all quantum correlations and that even correlations of separable states are
not completely classical. Entanglement is an inevitable feature of not only quantum
theory but also any non-classical theory~\cite{RichensSelbyAl-Safi}, and this is
necessary for emergent classicality in all physical theories. The study of quantum
correlation quantifiers other than entanglement, such as QD, has a crucial importance for
the full development of new quantum technologies because it is more robusted than
entanglement against the effects of decoherence~\cite{Werlang,Modi,Yune,Wang} and can be
among others a resource in quantum computation~\cite{Dattaa,Dattab,Ke}, quantum
non-locality~\cite{Liu}, quantum key distribution~\cite{Brodutch}, remote state
preparation~\cite{Dakic}, quantum cryptography~\cite{Pirandola} and quantum
coherence~\cite{Yao}.
\\
\par\indent%
The QD, as a quantum correlation of a bipartite system, initially introduced by Olliver
and Zurek~\cite{Zurek,Ollivier} and by Henderson and Vedral~\cite{Henderson}, is a more
general concept to measure quantum correlations than quantum entanglement, since
separable mixed states can have nonzero QD. This measures the fraction of the pairwise
mutual information that is locally inaccessible in a multipartite system. The QD is
defined as the difference between the total and classical correlations coded in the same
state, given by $\mathcal{I}_{AB}=S[\rho_A]+S[\rho_B]-S[\rho_{AB}]$  and
$\mathcal{J}_{AB}=S[\rho_B]-S_{X|\{\Pi^{Y}_m\}}(\rho_{AB})$, respectively, where
$X=\{A,B\}$ with $Y=\{B,A\}$, and $\Pi^{Y}_m$ is a measurement carried out in the
partition $A$ or $B$. These correlations are also known as quantum mutual information and
conditional mutual information, respectively. In this correlations,
$S[\rho]=-\tr{\rho\log_{2}\rho}=-\sum_{i}\lambda_{i}\log_{2}\lambda_{i}$ is the
von-Neumann entropy, where the $\lambda_{i}$'s are the eigenvalues of the density
operator $\rho$~\cite{Neumann}; the density operator $\rho_A=\tr[B]{\rho_{AB}}$
($\rho_B=\tr[A]{\rho_{AB}}$) is the reduced state in the partition $A$ ($B$) and
$S_{X|\{\Pi^{Y}_m\}}(\rho_{AB})$ is the conditional entropy of $X$ due to a measure in
$Y$. The classical correlations measured in the partition $A$ and $B$ are written as
$\mathcal{J}_{\overrightarrow{AB}}$ and $\mathcal{J}_{\overleftarrow{AB}}$, respectively.
The QD is also called the \emph{locally inaccessible information} (LII)~\cite{Fanchinib},
since $\delta_{\overrightarrow{AB}}=\mathcal{I}_{AB}-\mathcal{J}_{\overrightarrow{AB}}$
and $\delta_{\overleftarrow{AB}}=\mathcal{I}_{AB}-\mathcal{J}_{\overleftarrow{AB}}$ are
informations of the system that are inaccessible to an observer in the partition $A$ and
$B$, respectively. In this context, quantum measurements only provide information on the
partition measured, however, simultaneously they introduce disturbance and destroy the
coherence in the system. The QD or LII of any state $\rho_{AB}$, when performing measured
on the partition $A$, can be written as
\begin{subequations}\label{eq:43}%
\begin{eqnarray}
\hspace{-.4cm}\delta_{\overrightarrow{AB}}(\rho_{AB})&=&\mathcal{I}_{AB}(\rho_{AB})-\mathcal{J}_{\overrightarrow{AB}}(\rho_{AB})%
\nonumber\\%
\label{eq:43a}
&=&S[\rho_{A}]-S[\rho_{AB}]+S_{B|\{\Pi_m^A\}}(\rho_{AB}),
\end{eqnarray}
and when performing measured on the partition $B$ it is given by
\begin{eqnarray}
\hspace{-.4cm}\delta_{\overleftarrow{AB}}(\rho_{AB})&=&%
\mathcal{I}_{AB}(\rho_{AB})-\mathcal{J}_{\overleftarrow{AB}}(\rho_{AB})%
\nonumber\\%
\label{eq:43b}
&=&S[\rho_{B}]-S[\rho_{AB}]+S_{A|\{\Pi_m^B\}}(\rho_{AB}).
\end{eqnarray}
\end{subequations}
The difficult step is to find the conditional entropy because it requires a process of
minimization. The information in the unmeasured partition can be evaluated by the quantum
conditional entropy
\begin{subequations}\label{eq:44}%
\begin{eqnarray}
\label{eq:44a}
\hspace{-.5cm}S_{B|\{\Pi_m^A\}}\left(\rho_{AB}\right)%
&=&\min_{\{\Pi_m^A\}}\sum_{m}\vmedio{\Pi_m^A}_\rho \ S_{B|\Pi_m^A}\left(\rho_{AB}\right),
\\%
\label{eq:44b}
\hspace{-.5cm}S_{A|\{\Pi_m^B\}}\left(\rho_{AB}\right)%
&=&\min_{\{\Pi_m^B\}}\sum_{m}\vmedio{\Pi_m^B}_\rho \ S_{A|\Pi_m^B}\left(\rho_{AB}\right),
\end{eqnarray}
\end{subequations}
where $S_{B|\Pi_m^A}\left(\rho_{AB}\right)$ and $S_{A|\Pi_m^B}\left(\rho_{AB}\right)$ are
the von-Neumann entropy of the partition $B$ and $A$ of $\rho_{AB}$ obtained after the
projective measurements ${\Pi_m^A}$ or ${\Pi_m^B}$, respectively. The QD is not always
larger than the entanglement~\cite{Luob,Li}, and there is not clear evidence of the
relationship between entanglement and quantum discord, in general, since they seem to
capture different properties of the states.
\\
\par\indent%
Experimentally, it is difficult to prepare pure states. In general, the states are mixed since
they characterize the interaction of the system with its surrounding environment. The study of
the quantum information properties of mixed states is more complicated and lesser understood
than that of pure states. The set of Werner states~\cite{Wernera} is an important type of
mixed states, derived in 1989, which plays a fundamental role in the foundations of quantum
mechanic and quantum information theory. Since these states admit a hidden variable model
without violating Bell's inequalities, then the correlation measured that are generated with
these states can also be described by a local model, despite of being entangled. Moreover,
these states are used as quantum channels with noise that do not maintain the additivity, they
are also in the study of deterministic purifications (see references in~\cite{Lyons2012}).
\\
\par\indent%
It is important to clarify the main differences between the Werner states (Ws) and the
Generalized Werner-Like states (GWLs). The bipartite \emph{Werner states} of qubits are
self-adjoint operators, bounded and class trace that act onto the composite space
$\mathscr{H}_2\otimes\mathscr{H}_2$, where $\mathscr{H}_2$ is the Hilbert space of
dimension two, formed by an admixture convex of the exchange operator previously
normalized
\vspace{-.5cm}%
\begin{equation}\label{eq:51}
\tfrac{1}{2}\hat{\mathbb{F}}_4=\tfrac{1}{2}\sum\limits_{i,j=0}^{3}\ket{ij}\bra{ji},
\end{equation}
\mbox{}\\*[-.4cm]%
with the maximally mixed state, also is designated as white noise, given for
\vspace{-.5cm}%
\begin{equation}\label{eq:52}
\hat{\openone}_{4}=\tfrac{1}{4}\sum\limits_{i,j=0}^{3}\ket{ij}\bra{ij}.
\end{equation}
\mbox{}\\*[-.4cm]%
So that the Ws are written as
\begin{equation}\label{eq:0}%
\rho_{\textrm{W}}(p)=\tfrac{1-p}{4}\hat{\openone}_{4}+\tfrac{p}{2}\hat{\mathbb{F}}_4,%
\quad\textrm{where}\quad p\in\left[-1,\tfrac{1}{3}\right].
\end{equation}
The range of variation of the mixing parameter $p$ guarantees the positivity of \eqref{eq:0},
furthermore, Volberechet and Werner show in \cite{volbechet2001} that the EoF for the states
defined by the equation \eqref{eq:0} is
\begin{small}
\begin{equation}\label{eq:45}%
EoF_{\textrm{W}}(p)=\begin{cases}
0&\textrm{if\;}-\tfrac{1}{3}\le p\le\tfrac{1}{3}\\
H_2\left(\tfrac{2-\sqrt{4-\left(3p+1\right)^2}}{4}\right) &\textrm{if\;}-1\le p<-\tfrac{1}{3}%
\end{cases}
\end{equation}
\end{small}
\mbox{}\hspace{-.3cm} where $H_{2}(x)=-x\log_{2}(x)-(1-x)\log_{2}(1-x)$ is the Shannon
binary entropy function. The Ws, given in the equation \eqref{eq:0}, are invariant under
any unitary operator of the form $\hat{\mathbb{U}}\otimes\hat{\mathbb{U}}$ and admit a
model of hidden variables if $-1\le p\le-\tfrac{1}{2}$ (see \cite{Wernera}), being still
entangled. The Ws is pure only when $p=-\tfrac{1}{3}$, being the same as the Bell state
$\ket{\Phi_-}=(\ket{01}-\ket{10})/\sqrt{2}$, this is
\begin{equation}\label{eq:9}%
\rho_{\textrm{W}}\left(-\tfrac{1}{3}\right)=
\tfrac{1}{2}(\hat{\openone}_{4}-\hat{\mathbb{F}}_4)=\ket{\Phi_-}\bra{\Phi_-}.
\end{equation}
For all $p\neq-\tfrac{1}{3}$ the Ws are mixed.
\\
\par\indent%
On the other hand, the GWLs, \emph{Quasi-Werner states}~\cite{Horodecki-PhysRevA.59.4206}
or \emph{Werner-Popescu states}~\cite{Popescu-PRL.72.797}, for qubits, are a
one-parametric family of mixed states, being the sum convex between the maximally mixed
state \eqref{eq:52} and an pure state $\ket{\psi}$, where the density matrix of order $4$
for the GWLs has the form
\begin{equation}\label{eq:4}%
\rho_{\textrm{GWL}}(\psi,p)=\tfrac{1- p}{4}\hat{\openone}_{4}+p\hat{\mathbb{P}}_{\psi},%
\quad\textrm{where}\quad\hat{\mathbb{P}}_{\psi}=\ket{\psi}\bra{\psi}.
\end{equation}
The range of variation of the mixing parameter $p$ is in this case $-\tfrac{1}{3}\leq
p\leq1$, which guarantees the positivity of the GWLs, shown in equation \eqref{eq:4}.
Here the parameter $p$ is considered as a probability when the range of variation is
$0\le p\le1$, in this case equation \eqref{eq:4} represents a convex sum of a pure state
$\ket{\psi}$ and white noise \eqref{eq:52}, with probabilities $p$ and $1-p$,
respectively. The fundamental difference between the states \eqref{eq:0} and \eqref{eq:4}
is that $\hat{\mathbb{F}}$ is an involutive operator
($\hat{\mathbb{F}}^2=\hat{\openone}$) while $\hat{\mathbb{P}}_\psi$ is an idempotent
operator ($\hat{\mathbb{P}}^2_\psi=\hat{\mathbb{P}}_\psi$), generating different
correlations since the replacing of $\tfrac{1}{2}\hat{\mathbb{F}}$ by
$\hat{\mathbb{P}}_{\psi}$ in equation \eqref{eq:0} makes $\rho_{\textrm{W}}(p)$ not
unitarily equivalent to $\rho_{\textrm{GWL}}(\psi,p)$.
\\
\par\indent%
Otherwise, the Bell Werner-Like states (BWLs), also called \emph{noisy
singlets}~\cite{Horodeckia}, are obtained by using the Bell states
$\ket{\Psi_{\pm}}=\tfrac{1}{\sqrt{2}}(\ket{00}\pm\ket{11})$ and
$\ket{\Phi_{\pm}}=\tfrac{1}{\sqrt{2}}(\ket{01}\pm\ket{10})$ as projectors in equation
\eqref{eq:4}. This states are maximally entangled and have been studied widely as a
fundamental resource for the quantum information processing, and also in the study of
non-local properties in quantum mechanics. The noisy singlets and the Ws are connected by
the transformation
\begin{equation}\label{eq:53}
\rho_{\textrm{GWL}}(\Phi_{-},-p)=\tfrac{1+p}{4}\,\hat{\openone}_4-p\,\hat{\mathbb{P}}_{\Phi_{-}}
\equiv\rho_{\textrm{W}}.
\end{equation}
This equality is exact only in four dimensions. In other dimensions it is impossible obtain
this equality. The BWLs obtained from $\ket{\Psi_+}$ and $\ket{\Phi_-}$ have been employed in
the study of the QD ~\cite{Ollivier,Henderson}, and they can be considered as a particular
case of two-qubit $X-$states~\cite{Lia,Ali}. Nevertheless, any unitary transformation applied
on GWLs leaves them invariant in shape, without changing the mixing parameter, is
\begin{equation}\label{eq:5}%
\begin{split}
\rho_{\textrm{GWL}}(\psi,p)\xrightarrow{\hat{\mathbb{U}}}%
\hat{\mathbb{U}}\rho_{\textrm{GWL}}(\psi,p)\hat{\mathbb{U}}^\dag%
&=\tfrac{1-p}{4}\hat{\openone}_{4}+p\hat{\mathbb{P}}_{\psi_U}
\\%
&=\rho_{\textrm{GWL}}(\psi_U,p),
\end{split}
\end{equation}
where $\ket{\psi_U}=\hat{\mathbb{U}}\ket{\psi}$. The GWLs changed by unitary
transformations are called Werner derivative states~\cite{Hiroshima2000}, however, the
study described in reference~\cite{Hiroshima2000} is uncomplete since it only considers a
particular class of unitary transformations. In this article, we show that Werner
derivative states they have the same EoF and QD.
\\
\par\indent%
Until now, due that the quantitative evaluation of QD involves an optimization procedure over
all posible measurement on one of the subsystems under study the explicit expression of QD
only has been obtained for a few special classes of two-qubit
$X-$states~\cite{LuoPRA77.042303.2008,Ali,Maldonado,Yurischev}, and generally this is
determined numerically~\cite{Girolami}. The principal aim of this paper is to derive
analytical solutions of EoF and QD for the GWLs built with generalized pure states in
bipartite systems of qubits.
\\
\par\indent%
In order to illustrate the relevance of our results, we study the QD of the
GWLs associated to one bipartite entangled $f-$deformed (or nonlinear) coherent
state. In this context, Man'ko and collaborators~\cite{Mankoa} introduced the
$f-$deformed oscillators as a generalization of
$q-$oscillators~\cite{Biedenharn,Macfarlane}. These nonlinear coherent states
exhibit some nonclassical features such as quadrature
squeezing~\cite{Roknizadeh,Harounib}, second order squeezing~\cite{Tavassolyb},
sub-Poissonian~\cite{Roknizadeh,Harounib} and super-Poissonian
statistics~\cite{Recamierb,Dehdashti}, antibunching effect~\cite{Santos}, and
negativity of Wigner function in parts of the phase space~\cite{Tavassolyb}.
The $f-$deformed coherent states have been used: to evaluate the statistical
behavior of nonlinear coherent states associated to the Morse and
P\"{o}schl-Teller Hamiltonians~\cite{Recamierb}, to describe the center-of-mass
motion of a trapped ion~\cite{Aniello,Yazdanpanah,Harounia,Harty}, to study
quantum dot exciton states~\cite{Harounib}, the nonclassical properties of
deformed photon-added nonlinear coherent states~\cite{Safaeian} and
$f$-deformed intelligent states~\cite{Tavassoly}, to produce the superposition
of nonlinear coherent states and entangled coherent
states~\cite{Karimia,Karimib}, to describe non-linear coherent states by
photonic lattices~\cite{Dehdashti}, among other applications. In this work, we
studied the analytical results obtained for the QD and EoF associated to
bipartite Werner-Like $f-$deformed coherent states in the following cases: the
center-of-mass motion of trapped ions with P\"{o}schl-Teller potential, the
entangled exciton states in a quantum dot, and the entangled diatomic molecules
using the deformed Morse potential  function.
\\
\par\indent%
The paper is organized as follows. In Sec. II y IV we present an analytical approach to
obtain the exact solutions of EoF and the QD for the GWLs, while in Sec. III we present a
technique that allows obtaining exact solutions of the QD for the Ws. In Sec V we present
some applications, first, we consider the case of discrete states, where we illustrate
the monotonous behavior of the QD with the concurrence of a pure state. In this section,
also we present an algebraic review to the bipartite entangled $f-$deformed coherent
states when they are obtained as eigenstates of the deformed annihilation operator, as
well as when they are obtained by the application of the deformed displacement operator
on the vacuum state. Several deformation functions that we use later in the paper are
presented in this section. In this section, also is devoted to illustrate the behavior of
QD and EoF of bipartite entangled $f-$deformed GWLs. Finally in Sec. VI y VII we present
the analysis and conclusions drawn from our results.

\section{Entanglement of formation of Ws and GWLs}
\label{sec:EoFWL}%
A good measure to quantify the entanglement of a pure state $\ket{\psi}$ is the
von-Neumann entropy, since a pure state can be constructed from a set of maximally
entangled singlet states and the number of these states is proportional to the entropy of
the reduced states of any partitions~\cite{Woottersa,Woottersb}. So the EoF for a pure
state $\ket{\psi}$ is given by
\begin{equation}\label{eq:32}%
EoF_\psi=S[\rho_A]=S[\rho_B],
\end{equation}
where $\rho_X=\tr[X]{\ket{\psi}\bra{\psi}}$ with $X=A$ or $B$. It is clear that the EoF
does not change under local unitary operations, so it is not possible to create or
destroy entanglement using these transformations. However, the von-Neumann entropy is not
a good measure of the degree of entanglement for mixed states because there are product
states whose partitions may have entropies different from zero, for example,
$\rho=\rho_1\otimes\rho_2$ with $S[\rho_1]\neq0$. To quantify the degree of entanglement
for any states belonging to $\mathscr{H}_2\otimes\mathscr{H}_2$,
Wooters~\cite{Woottersa,Woottersb} proposed that the EoF for any states $\rho$ (pure or
mixing) is
\begin{equation}\label{eq:33}%
EoF_\rho=H_2\left(\tfrac{1+\sqrt{1-C^2[\rho]}}{2}\right),
\end{equation}
where $C[\rho]$ is the \emph{concurrence} function of the state $\rho$, defined as
$C[\rho]=\max\{0,\sqrt{\lambda_1}-\sqrt{\lambda_2}-\sqrt{\lambda_3}-\sqrt{\lambda_4}\}$.
The $\lambda_i$'s are the eigenvalues of the positive operator $\rho\widetilde{\rho}$,
arranged in decreasing order. The operator $\widetilde{\rho}$ is the spin-flip operation
on the conjugate of the state $\rho$, i.e.
$\widetilde{\rho}=(\sigma_y\otimes\sigma_y)\overline{\rho}(\sigma_y\otimes\sigma_y)$,
being $\overline{\rho}$ the conjugate complex of $\rho$. Here the difficult step is to
evaluate the concurrence of the state. In this section we find the eigenvalues of
$\rho\widetilde{\rho}$ for the GWLs.
\\%
\par\indent%
In the case of a pure state $\ket{\psi}$, the spin-flip operation onto the conjugate
complex of the state is given by
$\widetilde{\rho}=(\sigma_y\otimes\sigma_y)\ket{\overline{\psi}}\bra{\overline{\psi}}(\sigma_y\otimes\sigma_y)\equiv\ket{\widetilde{\psi}}\bra{\widetilde{\psi}}$
so
$\rho\widetilde{\rho}=\braket{\psi}{\widetilde{\psi}}\ket{\psi}\bra{\widetilde{\psi}}$,
and the characteristic equation $\rho\widetilde{\rho}\ket{\lambda}=\lambda\ket{\lambda}$
leads to $\lambda=|\braket{\psi}{\widetilde{\psi}}|^2$, after projecting this equation on
$\bra{\widetilde{\psi}}$. Also, the determinant of $\ket{\psi}\bra{\widetilde{\psi}}$ is
zero and therefore $\rho\widetilde{\rho}$ has a null eigenvalue with multiplicity three
which corresponds to the ortogonal projection to the state $\ket{\widetilde{\psi}}$. In
this sense, $\sqrt{\lambda_1}=|\braket{\psi}{\widetilde{\psi}}|$ and
$\lambda_2=\lambda_3=\lambda_4=0$, being the concurrence for a pure state $\ket{\psi}$
\begin{equation}\label{eq:34}%
C[\ket{\psi}]=|\braket{\psi}{\widetilde{\psi}}|=|\bra{\psi}\sigma_{y}\otimes\sigma_{y}\ket{\overline{\psi}}|.
\end{equation}
This result is known as the Wooters~\cite{Woottersa} formule  for pure states.
\\%
\par
\begin{figure}[t]
\begin{picture}(0,4.7)(4.7,.2)
\put(.5,-.2){%
\put(0,0){\includegraphics[height=4.5cm,width=8cm]{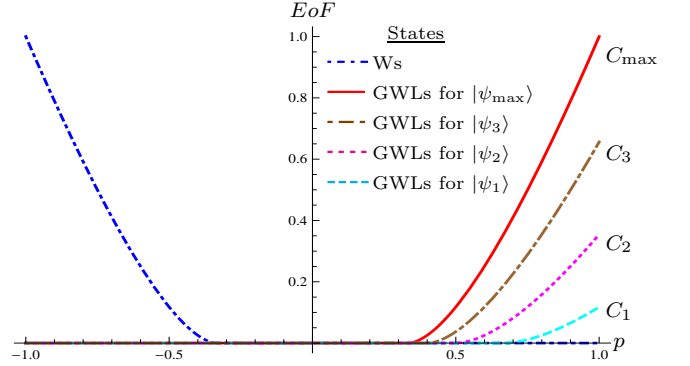}}%
\put(8,.2){\footnotesize$p$}\put(3.7,4.6){\footnotesize$EoF$}%
\put(7.9,.5){%
\put(0,3.5){\footnotesize$C_{\max}$}%
\put(0,2.2){\footnotesize$C_{3}$}%
\put(0,1){\footnotesize$C_{2}$}%
\put(0,.1){\footnotesize$C_{1}$}%
}%
\put(4.2,4){%
 \put(.8,.3){\scriptsize \underline{States}}
 \psline[linecolor=blue](0,0)(.05,0)\psline[linecolor=blue](.13,0)(.23,0)\psline[linecolor=blue](.31,0)(.36,0)\psline[linecolor=blue](.44,0)(.54,0)%
 \put(.6,-.1){\scriptsize  Ws}%
 \psline[linecolor=red](0,-.4)(.5,-.4)%
 \put(.6,-.5){\scriptsize  GWLs for $\ket{\psi_{\max}}$}%
 \psline[linecolor=brown](0,-.8)(.07,-.8)\psline[linecolor=brown](.15,-.8)(.35,-.8)\psline[linecolor=brown](.43,-.8)(.5,-.8)%
 \put(.6,-.9){\scriptsize  GWLs for $\ket{\psi_3}$}%
 \multiput(0,-1.2)(.15,0){4}{\psline[linecolor=magenta](0,0)(.07,0)}%
 \put(.6,-1.3){\scriptsize GWLs for $\ket{\psi_2}$}%
 \multiput(0,-1.6)(.15,0){4}{\psline[linecolor=cyan](0,0)(.1,0)}%
 \put(.6,-1.7){\scriptsize GWLs for $\ket{\psi_1}$}%
         }%
}%
\end{picture}
\caption{(color online) EoF of Ws and GWLs for the pure states $\ket{\psi_{\max}}$,
$\ket{\psi_3}$, $\ket{\psi_2}$ and $\ket{\psi_1}$ those concurrences are equals to
$C_{\max}=1$, $C_3=3/4$, $C_2=1/2$ and $C_1=1/4$, respectively. } \label{fig:1-EoF}
\end{figure}

\indent%
In the case of GWLs, the spin-flip operation applied on the conjugate complex of the
states defined by equation \eqref{eq:4} is given by
\begin{equation}\label{eq:35}%
\widetilde{\rho}_{\textrm{GWL}}(\psi,p)=%
\tfrac{1-p}{4}\hat{\openone}_4+p\ket{\widetilde{\psi}}\bra{\widetilde{\psi}}\equiv%
\rho_{\textrm{GWL}}(\tilde{\psi},p),
\end{equation}
while
\begin{equation}\label{eq:36}%
\rho_{\textrm{GWL}}(\psi,p)\widetilde{\rho}_{\textrm{GWL}}(\psi,p)=%
\left(\tfrac{1-p}{4}\right)^2\hat{\openone}_4+\hat{\mathbb{A}},%
\end{equation} where
\begin{equation}\label{eq:37}%
\hat{\mathbb{A}}=p^2C[\psi]\exp^{i\phi}\ket{\psi}\bra{\widetilde{\psi}}+%
\tfrac{p(1-p)}{4}\left(\ket{\psi}\bra{\psi}+\ket{\widetilde{\psi}}\bra{\widetilde{\psi}}\right).
\end{equation}
Here we have replaced $\braket{\psi}{\widetilde{\psi}}$ by $C[\psi]\exp^{i\phi}$, where
$\phi$ is the argument of $\braket{\psi}{\widetilde{\psi}}$. On the other hand, the
eigenvectors of the matrix $\hat{\mathbb{A}}$ are equal to the eigenvectors of
$\rho_{\textrm{GWL}}(\psi,p)\widetilde{\rho}_{\textrm{GWL}}(\psi,p)$, so we will focus on
finding the eigenvalues of this matrix. It is clear from equation \eqref{eq:37} that the
domain of $\hat{\mathbb{A}}$ is the linear capsule generated or expanded by
$\{\ket{\psi},\ket{\widetilde{\psi}}\}$, which means that the eigenvectors of
$\hat{\mathbb{A}}$ belong to this capsule and they can be written as
$\ket{\lambda}=\Lambda_1\ket{\psi}+\Lambda_2\ket{\widetilde{\psi}}$. Projecting the
equation $\hat{\mathbb{A}}\ket{\lambda}=\lambda\ket{\lambda}$ into the linear capsule, we
obtain an equation system for $\braket{\psi}{\lambda}$ and
$\braket{\tilde{\psi}}{\lambda}$ from which a straightforward calculation yields
\begin{equation}\label{eq:39}%
\begin{bmatrix}
\tfrac{p(1-p)}{4}-\lambda   & \frac{p(1+3p)}{4}C[\psi]\exp^{i\phi} \\*[.2cm]
\tfrac{p(1-p)}{4}C[\psi]\exp^{-i\phi} & \tfrac{p(1-p)}{4}+p^2C^2[\psi]-\lambda%
\end{bmatrix}\,
\begin{bmatrix}
\braket{\psi}{\lambda}
\\*[.2cm]%
\braket{\widetilde{\psi}}{\lambda}
\end{bmatrix}=0.
\end{equation}
To determine a solution other than the trivial one, we impose that the determinant of the equation system is zero and
obtain the following eigenvalue equation
\begin{equation}\label{eq:40}%
\lambda^2-\frac{p^2(1-2\Delta_0^2)+p}{2}\lambda+\frac{p^2(1-p)^2}{16}\Delta_0^2=0,
\end{equation}
where $\Delta_{0}\definido\sqrt{1-C^2[\ket{\psi}]}$. From this equation, two eigenvalues
are determined. The other two eigenvalues of the operator $\hat{\mathbb{A}}$ that
correspond to the eigenvector expanded into the linear capsule orthogonal to
$\{\ket{\psi},\ket{\widetilde{\psi}}\}$ are zero because $\det(\widehat{\mathbb{A}})=0$.
Finally, the eigenvalues of equation \eqref{eq:36} are in decreasing order
\begin{subequations}\label{eq:41}
\begin{eqnarray}
\label{eq:41a}%
&&\hspace{-1.1cm}\lambda_1\hspace{-.1cm}=\hspace{-.1cm}%
\left(\tfrac{1-p}{4}\right)^2\hspace{-.1cm}+%
\tfrac{p(1-p+2pC^2[\psi])+|p|C[\psi]\sqrt{(1+p)^2-4p^2\Delta_0^2}}{4},
\\%
\label{eq:41b}%
&&\hspace{-1.1cm}\lambda_2\hspace{-.1cm}=\hspace{-.1cm}%
\left(\tfrac{1-p}{4}\right)^2\hspace{-.1cm}+%
\tfrac{p(1-p+2pC^2[\psi])-|p|C[\psi]\sqrt{(1+p)^2-4p^2\Delta_0^2}}{4},
\end{eqnarray}
and
\begin{eqnarray}
\label{eq:41c}%
&&\hspace{-1cm}\lambda_3=\lambda_4=\left(\tfrac{1-p}{4}\right)^2,
\end{eqnarray}
\end{subequations}
so that the concurrence for GWLs is given by
\begin{equation}\label{eq:42}%
C[\rho_{\textrm{GWL}}]=\max\left\{0,\sqrt{\lambda_1}-\sqrt{\lambda_2}-\tfrac{1-p}{2}\right\}.
\end{equation}
This shows that GWLs are separable when $-\tfrac{1}{3}\le p\le\tfrac{1}{1+2C[\psi]}$ and
entangled when $\tfrac{1}{1+2C[\psi]}<p\le1$. In particular, for BWLs we have the usual
result~\cite{Horodecki-PhysRevA.59.4206}, i.e., they are entangled if $1/3<p\le1$ and
classically correlated if $-1/3\le p\le1/3$, since $\ket{\psi}$ is maximally entangled, i.e.
$C[\psi]=1$. When the pure state $\ket{\psi}$ is a product state ($C[\psi]=0$) then all the
GWLs are a convex sum of product states. In Fig.~\ref{fig:1-EoF} is shown the EoF as a
function of the mixing parameter $p$, for Ws and GWLs associated to the pure states
$\ket{\psi_1}$, $\ket{\psi_2}$, $\ket{\psi_3}$ and $\ket{\psi_{\max}}$, with the values of
concurrence $C_1=1/4$, $C_2=1/2$, $C_3=3/4$ and $C_{\max}=1$, respectively. The EoF of Ws is
given by equation \eqref{eq:45} We observe that the EoF of the GWLs increase with the
concurrence of the pure state asociated to the GWLs. The BWLs are the ones that have the
maximum entanglement. The Ws and GWLs are entangled in diferent regions, and the maximum value
of $p$, where the EoF is zero, for the pure states $\ket{\psi_1}$, $\ket{\psi_2}$,
$\ket{\psi_3}$ and $\ket{\psi_{\max}}$ are given by $2/3$, $1/2$, $2/5$ and $1/3$,
respectively.
\section{Quantum discord of Werner states}
\label{sec:QDW}%
The fundamental amount for the study of quantum information is the von-Neumann entropy,
namely, the information in terms of its uncertainty. This quantity measures the expected value
of quantum information content~\cite{wilde2011}. For pure states, the von-Neumann entropy is
zero, because the density operator is a projector of rank one, and represent full knowledge
about the state of quantum system. For the $4-$dimensional Hilbert space
$\mathscr{H}_2\otimes\mathscr{H}_2$ maximal uncertainty is represented by the completely mixed
density operator \eqref{eq:52}, with a value for the von-Neumann entropy of $2$, in bits. In
these systems one has $0\leq S[\rho]\leq2$. Thus, the entropy for the Ws given in \eqref{eq:0}
will be bounded between these two values, being zero when $p=-1$ and maximum when $p=0$.
\\
\par\indent%
The starting point is to obtain the entropy of the Ws to find the eigenvalues of the states
given under the equation \eqref{eq:0}. Is easily demonstrated that the eigenvalues del
exchange operator are $\pm1$, since\vspace{-.1cm}
\begin{gather}
\hat{\mathbb{F}}_4\ket{f}=f\ket{f}%
\;\Longrightarrow\;%
\hat{\mathbb{F}}_4^2\ket{f}=f\hat{\mathbb{F}}_4\ket{f}%
\;\Longrightarrow\;%
\hat{\openone}_4\ket{f}=f^2\ket{f}
\nonumber\\
\label{eq:54}%
\;\Longrightarrow\;%
\ket{f}=f^2\ket{f}\quad\therefore\quad f=\pm1.
\end{gather}
However, the exchange operator $\hat{\mathbb{F}}_4$ acts onto
$\mathscr{H}_2\otimes\mathscr{H}_2$ so it must has $4$ eigenvalues, which are not simple. Let
$\ket{\lambda_{\textrm{W}}^\pm}$ be the eigenvectors of the exchange operator (that are same
to eigenstates of the werner states) with  eigenvalues $1$ and $-1$, respectively, then,
\begin{subequations}\label{eq:56}
\begin{align}
\label{eq:56a}%
\hat{\mathbb{F}}_4\ket{\lambda_{\textrm{W}}^+}&=+\ket{\lambda_{\textrm{W}}^+}%
\;\Longrightarrow\;%
\braket{ji}{\lambda_{\textrm{W}}^+}=\braket{ij}{\lambda_{\textrm{W}}^+},
\\
\label{eq:56b}%
\hat{\mathbb{F}}_4\ket{\lambda_{\textrm{W}}^-}&=-\ket{\lambda_{\textrm{W}}^-}%
\;\Longrightarrow\;%
\braket{ji}{\lambda_{\textrm{W}}^-}=-\braket{ij}{\lambda_{\textrm{W}}^-}.
\end{align}
\end{subequations}
This proves that the eigenstates $\ket{\lambda_{\textrm{W}}^+}$ and
$\ket{\lambda_{\textrm{W}}^-}$ belong to spaces of dimensions $\frac{2(2+1)}{2}=3$ and
$\frac{2(2-1)}{2}=1$, respectively. The multiplicity of this eigenvalues is the dimensions of
those spaces. The Ws has one simple eigenvalue, given for $\frac{1-3p}{4}$, and $3$
degenerates eigenvalues with value $\frac{1+p}{4}$. This allows to obtain the von-Neumann
entropy for the Ws, given for,
\begin{equation}\label{eq:55}
\begin{split}
&S_{AB}(p)\definido S[\rho_{\textrm{W}}]%
=-\tr{\rho_{\textrm{W}}\log_{2}\rho_{\textrm{W}}}
\\
&=2-\tfrac{1-3p}{4}\log_{2}(1-3p)-\tfrac{3(1+p)}{4}\log_{2}(1+p).
\end{split}
\end{equation}
This expression is a concave function of the mixing parameter, being zero for $p=-1$ and
two for $p=0$.
\\
\par\indent%
The partial trace of the exchange operator in any partition is equal to identity operator in
two dimentions, this is, $\textrm{tr}_{X}[\hat{\mathbb{F}}_4]=\hat{\openone}_2$ with $X=A$ or
$X=B$. This allows determine the quantum information of each partition of the system
$\mathscr{H}_2\otimes\mathscr{H}_2$ contained in the Ws. Therefore, the reduced state of the
Ws is a maximally mixed state and the entropy of those states is the logarithm of the
dimension of the reduced space, this is, for the state
\begin{subequations}\label{eq:48}
\begin{gather}
\label{eq:48a}%
\rho_{\textrm{W}}^X=\tr[Y]{\rho_{\textrm{W}}}=\tfrac{1}{2}\openone_2%
\intertext{we have,}%
\label{eq:48b}%
S_X(p)=-\tr{\rho_{\textrm{W}}^X\log_2\rho_{\textrm{W}}^X}=1.
\end{gather}
\end{subequations}
Where $X=\{A,B\}$ when $Y=\{B,A\}$ in the equation \eqref{eq:48a}.
\\
\par\indent%
In order to quantify the QD, the entropy condicional is required, for this we perform a
projective measurement $\{\widehat{\Pi}_{m}\}$ on one partition of the subsystem. In the
partition $A$ we have
\begin{equation}\label{eq:8}
\widehat{\Pi}_{m}^{A}=\widehat{\Pi}_{m}\otimes\hat{\openone}_{2}=%
\tfrac{1}{2}\left[\hat{\openone}_{2}+(-1)^{m}\hat{n}\cdot\vec{\sigma}\right]\otimes\hat{\openone}_{2},
\end{equation}
with $m=0$ or $m=1$. Here
$\hat{n}=\sin(2\theta)\cos(\phi)\vi+\sin(2\theta)\sin(\phi)\vj+\cos(2\theta)\vk$, is a
unitary vector on the Bloch sphere, and
$\vec{\sigma}=\sigma_{x}\vi+\sigma_{y}\vj+\sigma_{z}\vk$ is the Pauli vector. After a
local measurement $\widehat{\Pi}_{m}^{A}$, on the density matrix $\rho_{\textrm{W}}$, the
state of the system becomes a hybrid quasi-classical state~\cite{Modi}, this is, applying
the L\"{u}der rule~\cite{luders} to the Ws is obtained the post-measurement states
$\rho_{\textrm{W}|\Pi_{m}^{A}}$ (see appendix~\ref{sec:apenA})
\begin{subequations}\label{eq:16}
\begin{gather}
\label{eq:16a}%
\rho_{\textrm{W}}(p)\xrightarrow{\hspace{.2cm}\widehat{\Pi}_{m}^{A}\hspace{.2cm}}\rho_{\textrm{W}|\Pi_{m}^{A}}=
\frac{(\widehat{\Pi}_{m}^{A})\rho_{\textrm{W}}(\widehat{\Pi}_{m}^{A})^{\dag}}{p_{m}^{A}},
\\
\label{eq:16b}%
\rho_{\textrm{W}|\Pi_{m}^{A}}=\widehat{\Pi}_{m}\otimes\left\{\tfrac{1-p}{2}\hat{\openone}_2+p\widehat{\Pi}_m\right\},
\end{gather}
\end{subequations}
Where $p_m^A$ corresponds to the probability of reaching the state measured, which can be
evaluate as
\begin{equation}\label{eq:38}
p_m^A=\vmedio{\widehat{\Pi}_{m}^{A}}_{\rho_{\textrm{W}}}=\tr{\widehat{\Pi}_{m}^{A}\rho_{\textrm{W}}}=%
\tfrac{1}{2},
\end{equation}
The states in the partition $B$ of the equation \eqref{eq:16b} have the form of a GWLs in
$\mathscr{H}_2$ with mixing parameter $p$. The eigenvalues of these states are $\tfrac{1\pm
p}{2}$, and they are independent of the measure. For this reason, the conditional entropy is
same that the entropy of the reduced states in the partition $B$ of equation \eqref{eq:16b}.
Using \eqref{eq:44b} we obtain
\begin{small}
\begin{equation}\label{eq:49}%
S_{A|\{\Pi_m^B\}}(p)=\min_{\{\Pi_m^A\}}\sum_{m}p_m^A
S_{B|\Pi_m^A}\left(\rho_{\textrm{W}}\right)=H_2\left(\tfrac{1+p}{2}\right).%
\end{equation}
\end{small}A straightforward calculator shows that the conditional entropy in both
partitions are the same. Therefore, the QD is symmetric and it's given by
\begin{eqnarray}
\delta_{AB}(p)&=&\delta_{\overleftarrow{AB}}(p)=\delta_{\overrightarrow{AB}}(p)%
\nonumber\\
\label{eq:50}%
&&\hspace{-.5cm}\begin{split}%
=&\; H_2\left(\tfrac{1+p}{2}\right)-1+\tfrac{1-3p}{4}\log_{2}(1-3p)\\
&+\tfrac{3(1+p)}{4}\log_2(1+p).
\end{split}
\end{eqnarray}
\begin{figure}[t]
\begin{picture}(0,4.5)(3.5,0)
\put(-.8,-.2){%
\put(0,0){\includegraphics[height=4.5cm,width=8cm]{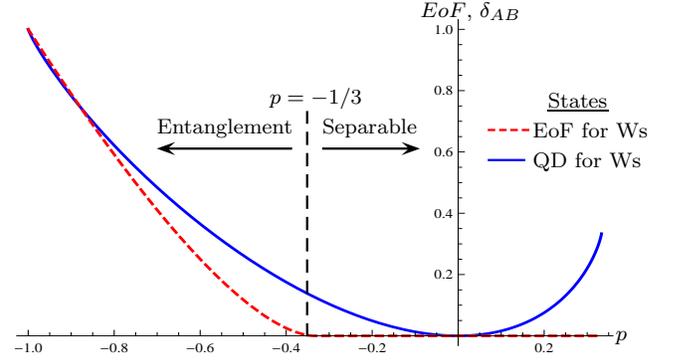}}%
\put(8,.2){\footnotesize$p$}\put(5.4,4.5){\footnotesize$EoF$, $\delta_{AB}$}%
\put(6.3,3){%
 \put(.8,0.3){\footnotesize\underline{States}}
 \multiput(0,0)(.15,0){4}{\psline[linecolor=red](0,0)(.1,0)}%
 \put(.6,-.1){\footnotesize EoF for Ws}%
 \psline[linecolor=blue](0,-.4)(.5,-.4)%
 \put(.6,-.5){\footnotesize QD for Ws}%
 }%
\put(3.9,.25){\psline[linestyle=dashed](0,0)(0,3)
          \psline{->}(.2,2.5)(1.5,2.5)
          \put(.2,2.7){\footnotesize Separable}%
          \psline{<-}(-2,2.5)(-.2,2.5)
          \put(-2,2.7){\footnotesize Entanglement}%
          \put(-.5,3.1){\footnotesize$p=-1/3$}
          }%
}%
\end{picture}
\caption{(color online) Plot of the QD (line blue solid) and EoF (line red dashed) for
the Ws} \label{fig:EoFandQDforWs}
\end{figure}
\par\indent%
In Fig.~\ref{fig:EoFandQDforWs} it shown the graph of QD and EoF for Ws. It is noted that
there is correlations, although the states not are entanglement. For $-1\le p<-0.88$ the EoF
is more big that the QD, in the rest of interval the relation it is reversed. The quantum
discord is only zero only for $p=0$ and for values of mixing parameter near of the origin is
very small.
\section{Quantum discord of GWLs}
\label{sec:QDWL}%
Firstly we evaluate the von-Neumann entropy. The GWLs given in the equation \eqref{eq:4}, has
a simple eigenvalue given by $\tfrac{1+3p}{4}$, and three degenerates eigenvalues with value
$\frac{1-p}{4}$, this allows to obtain the von-Neumann entropy, given by
\begin{equation}\label{eq:6}
\begin{split}
&S_{AB}[\psi,p]=S[\rho_{\textrm{GWL}}]%
=-\tr{\rho_{\textrm{GWL}}\log_{2}\rho_{\textrm{GWL}}}
\\
&\hspace{.5cm}=2-\tfrac{3(1-p)}{4}\log_{2} (1-p) - \tfrac{1+3p}{4}\log_{2} (1+3p).
\end{split}
\end{equation}
This expression is independent of the pure state $\ket{\psi}$, in addition to being a
monotonic function of the mixing parameter. Is clear from \eqref{eq:6} that the
information provided by the GWLs is minimal (maximum entropy) when $p=0$, corresponding
to a maximally mixed quantum state (white noise), while that the information is maximal
(minimum entropy) when $p=1$, value for which \eqref{eq:4} is pure.
\\
\par\indent%
To determine the quantum information (in term of its uncertainty) of each partition of the
system $\mathscr{H}_2\otimes\mathscr{H}_2$ contained in GWLs, given the equation \eqref{eq:4},
their is sufficient to take its partial traces, so that we have
\begin{subequations}\label{eq:7}
\begin{eqnarray}
\label{eq:7a}
\hspace{-.5cm}\rho_{\textrm{GWL}}^A&=&\tr[B]{\rho_{\textrm{GWL}}}=%
\tfrac{1-p}{2}\hat{\openone}_2+p\widehat{\mathbb{W}}_\psi\widehat{\mathbb{W}}_\psi^\dag,
\\
\label{eq:7b}
\hspace{-.5cm}\rho_{\textrm{GWL}}^B&=&\tr[A]{\rho_{\textrm{GWL}}}=%
\tfrac{1-p}{2}\hat{\openone}_2+p(\widehat{\mathbb{W}}_\psi^T)(\widehat{\mathbb{W}}_\psi^T)^\dag.
\end{eqnarray}
\end{subequations}
Where $\widehat{\mathbb{W}}_\psi\definido[\psi_{ij}]_{2\times2}$ is the matrix
constructed with the components of the pure state $\ket{\psi}$ in the computational base
$\ket{ij}$, namely $\psi_{ij}=\braket{ij}{\psi}$, while $\widehat{\mathbb{W}}^{T}_\psi$
is the transposed matrix of $\widehat{\mathbb{W}}_\psi$. The normalization condition of
the state $\ket{\psi}$ in term of matrix $\widehat{\mathbb{W}}_\psi$ is
$\tr{\widehat{\mathbb{W}}_\psi\widehat{\mathbb{W}}_\psi^\dag}=1$ (see
appendix~\ref{sec:apenA} for details). In this context the transpose connects the density
operator of both partitions and this operation does not modify the eigenvalues of the
reduce states. For this reason, the expressions \eqref{eq:7} show that the entropies of
the reduced states are equal, so that
\begin{equation}\label{eq:10}%
\begin{split}%
-\tr{\rho_{\textrm{GWL}}^A\log_2\rho_{\textrm{GWL}}^A}&\equiv-\tr{\rho_{\textrm{GWL}}^B\log_2\rho_{\textrm{GWL}}^B},\\
 S_A(\psi,p)&\equiv S_B(\psi,p).
\end{split}
\end{equation}
The two eigenvalues of $\widehat{\mathbb{W}}_\psi\widehat{\mathbb{W}}_\psi^\dag$ are
$\tfrac{1}{2}\left(1\pm\Delta_0\right)$ with $\Delta_0=\sqrt{1-C^2[\psi]}$, where
$C[\psi]$ is the concurrence of the pure state $\ket{\psi}$, belonging to
$\mathscr{H}_2\otimes\mathscr{H}_2$. This quantity is given by the formula of Wooters
\eqref{eq:34}
\begin{equation}\label{eq:11}%
C[\psi]=
2|\psi_{00}\psi_{11}-\psi_{01}\psi_{10}|\equiv%
2|\det\widehat{\mathbb{W}}_\psi|.
\end{equation}
These results show that the two eigenvalues of the reduced states \eqref{eq:7} are
$\tfrac{1}{2}\left(1\pm p\Delta_0\right)$, so the entropy \eqref{eq:10} takes the
following form
\begin{equation}\label{eq:12}
S_{A}(\psi,p)=S_{B}(\psi,p)=H_{2}\left(\tfrac{1+p\Delta_0}{2}\right).
\end{equation}
When $p=1$ in the equation \eqref{eq:12} one has the EoF given in the equation \eqref{eq:33}
of pure state $\ket{\psi}$; on the other hand, when $p=0$ the entropy of the reduced state is
maximal, take the value of one, which corresponds to a maximally mixed state in
$\mathscr{H}_2$.
\\
\par\indent%
In order to quantify the QD, the entropy condicional is required. We performed a projective
measurement \eqref{eq:8} on one partition $A$ of the subsystem. After of this local
measurement on the density matrix $\rho_{GWL}$, the state of the system becomes a hybrid
quasi-classical state~\cite{Modi}, this is, using the L\"{u}der rule~\cite{luders} we obtain
(see appendix~\ref{sec:apenA})
\begin{subequations}\label{eq:14}
\begin{gather}
\label{eq:14a}%
\rho_{\textrm{GWL}}\xrightarrow{\hspace{.3cm}\widehat{\Pi}_{m}^{A}\hspace{.3cm}}\rho_{\textrm{GWL}|\Pi_{m}^{A}}=
\frac{(\widehat{\Pi}_{m}^{A})\rho_{\textrm{GWL}}(\widehat{\Pi}_{m}^{A})^{\dag}}{p_{m}^{A}},
\\
\label{eq:14b}%
\rho_{\textrm{GWL}|\Pi_{m}^{A}}=\widehat{\Pi}_{m}\otimes\left\{\tfrac{1-x_m(p)}{2}\hat{\openone}_2+x_m(p)\ket{\widehat{\psi}}\bra{\widehat{\psi}}\right\},
\end{gather}
\end{subequations}
where $\rho_{GWL|\Pi_{m}^{A}}$ is the post-measurement state and $p_{m}^{A}$ corresponds
to the probability of reaching that state, which can be evaluated as
\begin{equation}\label{eq:15}%
p_{m}^{A}=\vmedio{\widehat{\Pi}_{m}^{A}}_{\rho_{\textrm{GWL}}}=\tr{\widehat{\Pi}_{m}^{A}\rho_{\textrm{GWL}}}=%
\tfrac{1-p}{2}+p\vmedio{\widehat{\Pi}_m^A}_\psi,
\end{equation}
with
\begin{equation}\label{eq:17}%
\vmedio{\widehat{\Pi}_m^A}_\psi=\tr{\widehat{\mathbb{W}}_\psi^{\dag}\widehat{\Pi}_{m}\widehat{\mathbb{W}}_\psi}.
\end{equation}
The amount $x_{m}(p)$ shown in \eqref{eq:16} is equivalente to a new mixing parameter of
the GWLs in the partition $B$, which given by (see appendix~\ref{sec:apenA})
\begin{equation}\label{eq:22}%
x_{m}(p)=\frac{p\langle\widehat{\Pi}_{m}^{A}\rangle_{\psi}}{\frac{1-p}{2}+p\langle\widehat{\Pi}_{m}^{A}\rangle_{\psi}}.%
\end{equation}
Noteworthy is that $x_{m}(p)$ is an injective function of the mixing parameters $p$, so both
parameters $p$ and $x_{m}(p)$ present the same variation range. It is important to see that
all $x_{m}(p)$ are not independent, since the sum over all probabilities
($\sum_{m}\langle\widehat{\Pi}_{m}^{A}\rangle_{\psi}=1$) impose a restriction on the $x_m(p)$,
given by
\begin{equation}\label{eq:18}%
\sum_m\frac{x_m(p)}{1-x_m(p)}=\frac{2p}{1-p}.
\end{equation}
In order to simplify the result \eqref{eq:16} we define the projector
$\ket{\widehat{\psi}}\bra{\widehat{\psi}}$ as
\begin{equation}\label{eq:19}%
\ket{\widehat{\psi}}\bra{\widehat{\psi}}=\sum_{i,j}%
\tfrac{\bra{i}{\widehat{\mathbb{W}}}_\psi^{\dag}\left(\hat{\openone}_{2}+(-1)^{m}\hat{n}\cdot\vec{\sigma}\right)
\widehat{\mathbb{W}}_\psi\ket{j}}{\tr{{\widehat{\mathbb{W}}}_\psi^{\dag}\left(\hat{\openone}_{2}+(-1)^{m}\hat{n}\cdot\vec{\sigma}\right)\widehat{\mathbb{W}}_\psi}}\,
\ket{j}\bra{i}.
\end{equation}
Thus, a projective measurement on the subsystem $A$ projects the system into a
statistical ensemble $\left\{p_{m}^{A},\rho_{AB|\Pi_{m}^{A}}\right\}$ quantifies the
information in the unmeasured partition as the quantum conditional entropy, given by
\begin{equation}\label{eq:20}
\begin{split}
S_{B|\{\Pi_m^A\}}(\psi,p)%
&=\min_{\{\Pi_m^A\}}\sum_{m}p_m^AS_{B|\Pi_m^A}(\psi,p),\\
=\tfrac{1}{2}\min_{\{\Pi_m^A\}}&\sum_{m}\tfrac{1-p}{1-x_m(p)} H_2\left(\tfrac{1+x_m(p)}{2}\right).
\end{split}
\end{equation}
Here the probability $p_m^A$ is replaced by the expression \eqref{eq:15}, while the
probability $\vmedio{\widehat{\Pi}_m^A}_\psi$ is written in terms of the mixing parameter
$x_m(p)$ using \eqref{eq:22}. $S_{B|\Pi_m^A}(\psi,p)$ is the von-Neumann entropy of the
partition $B$ of $\rho_{GWL}(\psi,p)$ obtained after the projective measurements
$\{\widehat{\Pi}_m^A\}$. Since the measurement might gives different results depending on the
basis choice, a minimization is taken over all possible rank-1 measurement
$\{\widehat{\Pi}_m^A\}$, applied on the subsystem $B$. Minimizing chooses the measurement of
$A$ that extracts as much information as possible of $B$. The hard step in the evaluation
quantum conditional entropy is usually the optimization of the conditional entropy
$S_{B|\Pi_{m}^{A}}$ over all projective measurements. However, it is clear that the process of
minimizing the conditional entropy is inceject to find the values of $x_{m}(p)$ that minimize
the probability $\vmedio{\widehat{\Pi}^A_m}_\psi$. In the Appendix~\ref{sec:apenB}, we show
that the conditional entropy of the partition $B$, have the form
\begin{equation}\label{eq:21}
S_{B|\{\Pi_m^A\}}(\psi,p)=F_p(\underline{x}_{0})+F_p(\underline{x}_{1}).
\end{equation}
where
\begin{equation}\label{eq:23}
F_p(x)=\tfrac{1-p}{2(1-x)}H_2\left(\tfrac{1+x}{2}\right)
\end{equation}
and the values $\left\{\underline{x}_{0},\underline{x}_{1}\right\}$ are such they
minimize the conditional entropy, for which $\underline{x}_{0}$ minimizes $F_p(x)$ but
$\underline{x}_{1}$ maximizes it (see appendix~\ref{sec:apenB}). In fact,
$\underline{x}_{0}$ is obtained when the probability $\vmedio{\widehat{\Pi}^A_m}_\psi$ is
minimized, while $\underline{x}_{1}$ is obtained by \eqref{eq:18}, so
\begin{subequations}\label{eq:24}%
\begin{equation}\label{eq:24a}%
\underline{x}_{0}=
\frac{p\vmedio{\widehat{\Pi}_{0}^{A}}^{\min}_{\psi}}{\frac{1-p}{2}+p\vmedio{\widehat{\Pi}_{0}^{A}}^{\min}_{\psi}}=
\frac{p(1-2A)}{1-2 pA},
\end{equation}
and
\begin{equation}\label{eq:24b}%
\underline{x}_{1}=\frac{2p-(1+p)x_0(p)}{1+p-2x_0(p)}=\frac{p(1+2A)}{1+2pA}.
\end{equation}
\end{subequations}
Can also be written as
\begin{equation}\label{eq:57}%
\underline{x}_{1}%
=\frac{p\vmedio{\widehat{\Pi}_{1}^{A}}^{\max}_{\psi}}{\frac{1-p}{2}+p\vmedio{\widehat{\Pi}_{1}^{A}}^{\max}_{\psi}}
=\frac{p(1-\vmedio{\widehat{\Pi}_{0}^{A}}^{\min}_\psi)}{\frac{1-p}{2}+p(1-\vmedio{\widehat{\Pi}_{0}^{A}}^{\min}_\psi)}\,.
\end{equation}
which also is consistent with \eqref{eq:22}. The value of $A$ showed in \eqref{eq:24}
(see appendix \ref{sec:apenB}) is given by
\begin{equation}\label{eq:25}%
A=\frac{1}{2}\sqrt{\sum_{i=1}^{3}\left(\tr{\widehat{\mathbb{W}}^{\dag}_\psi\sigma_{i}\widehat{\mathbb{W}}_\psi}\right)^2}.
\end{equation}
The equation \eqref{eq:21} is an analytical expression for the conditional entropy after a
measurement in partition $A$. The aforementioned procedure can be applied to obtain the
conditional entropy $S_{A|\{\Pi_m^B\}}(\psi,p)$, after a measurement in partition $B$. The
same result is obtained, except that instead of the matrix $\mathbb{\mathbb{W}}_\psi$, its
transpose is used, namely,
\begin{equation}\label{eq:26}
S_{A|\{\Pi_m^B\}}(\psi,p)=F_p(\underline{y}_{0})+F_p(\underline{y}_{1}) \ ,
\end{equation}
with
\begin{equation}\label{eq:28}%
\underline{y}_{0}=\frac{p(1-2B)}{1-2pB}%
\quad\textrm{and}\quad%
\underline{y}_{1}=\frac{p(1+2B)}{1+2pB}\,,
\end{equation}
and
\begin{equation}\label{eq:29}%
B=\frac{1}{2}\sqrt{\sum_{i=1}^{3}\left(\tr{(\widehat{\mathbb{W}}_\psi^T)^{\dag}\sigma_{i}\widehat{\mathbb{W}}_\psi^T}\right)^2}.
\end{equation}
Generally the QD is asymmetric and
$\delta_{\overrightarrow{AB}}\neq\delta_{\overleftarrow{AB}}$. We can study the average of
LII, defined as
$\varpi_{A|B}^{+}=(\delta_{\overrightarrow{AB}}+\delta_{\overleftarrow{AB}})/2$, and the
balance of LII, defined as
$\varpi_{A|B}^{-}=(\delta_{\overrightarrow{AB}}-\delta_{\overleftarrow{AB}})/2$ (see
reference~\cite{Fanchinia}). Nevertheless, a straightforward calculation showed that
\eqref{eq:25} and \eqref{eq:29} coincide, with which the QD in both partitions are iqual,
therefore the balance is zero. If we take the explicit forms of the entropy given in the
Eqs.~\eqref{eq:6}, \eqref{eq:12} and \eqref{eq:21}, we can obtain the analytical expressions
of the QD $\delta_{\overleftarrow{AB}}(\psi,p)$ and $\delta_{\overrightarrow{AB}}(\psi,p)$ for
the GWLs, after a measurement in partition $A$ or $B$. The exact analitical solutions are
\begin{eqnarray}
\delta_{AB}(\psi,p)&=&\delta_{\overleftarrow{AB}}(\psi,p)=\delta_{\overrightarrow{AB}}(\psi,p)%
\nonumber\\
&=&H_{2}\left(\tfrac{1-p\Delta_{0}}{2}\right)+F_p(\underline{x}_{0})+F_p(\underline{x}_{1})+
\nonumber\\%
\label{eq:30}
&+&\log_2\left(\tfrac{1-p}{4}\right)+\tfrac{1+3p}{4}\log_{2}\left(\tfrac{1+3p}{1-p}\right).
\end{eqnarray}
with $\Delta_0=\sqrt{1-C^2[\psi]}$, being $C[\psi]$ the concurrence of the pure state
$\ket{\psi}$ associated with the GWLs. The QD given in the equation~\eqref{eq:30}, in
addition to being symmetrical is a monotonous function of the concurrence $C[\psi]$. So,
all the pure states with the same concurrence $C[\psi]$ have equal QD, in the same way as
the EoF of the GWLs; this forms classes of equivalence among pure states $\ket{\psi}$
with equal concurrence.
\begin{figure}[t]
\begin{picture}(0,4.5)(3.5,0)
\put(-.8,0){%
\put(0,0){\includegraphics[height=4.5cm,width=8cm]{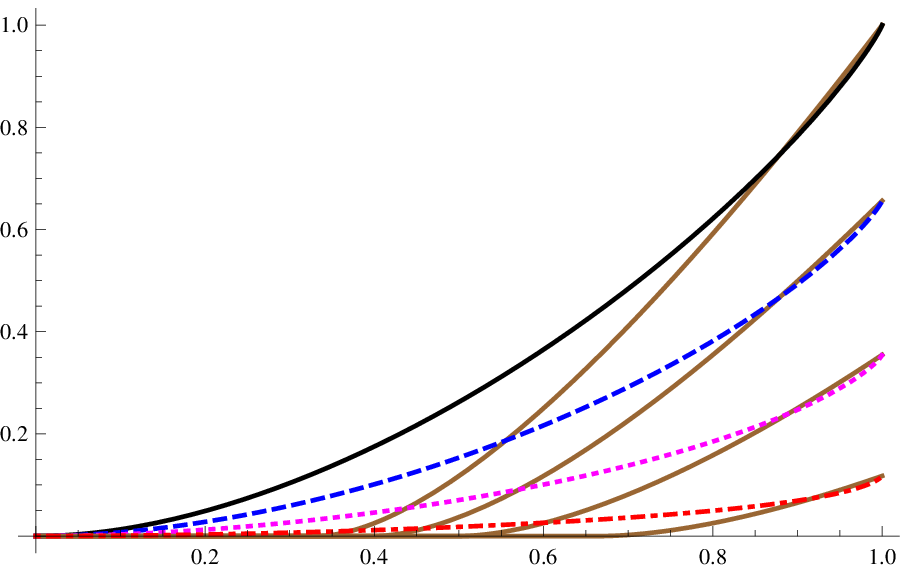}}%
\put(8,.2){\footnotesize$p$}\put(-.1,4.6){\footnotesize$EoF$, $\delta_{AB}$}%
\put(7.9,.5){%
\put(0,3.5){\footnotesize$C_{\max}$}%
\put(0,2.2){\footnotesize$C_{3}$}%
\put(0,1){\footnotesize$C_{2}$}%
\put(0,.1){\footnotesize$C_{1}$}%
}%
\put(.6,3.5){%
 \put(.8,0.3){\footnotesize\underline{States}}
 \psline[linecolor=brown](0,0)(.5,0)%
 \put(.6,-.1){\footnotesize EoF for GWLs}%
 \psline[linecolor=black](0,-.4)(.5,-.4)%
 \put(.6,-.5){\footnotesize QD for GWLs $\ket{\psi_{\max}}$}%
 \multiput(0,-.8)(.15,0){4}{\psline[linecolor=blue](0,0)(.1,0)}%
 \put(.6,-.9){\footnotesize QD for GWLs $\ket{\psi_3}$}%
 \multiput(0,-1.2)(.15,0){4}{\psline[linecolor=magenta](0,0)(.07,0)}%
 \put(.6,-1.3){\footnotesize QD for GWLs $\ket{\psi_2}$}%
 \psline[linecolor=red](0,-1.6)(.07,-1.6)\psline[linecolor=red](.15,-1.6)(.35,-1.6)\psline[linecolor=red](.43,-1.6)(.5,-1.6)%
 \put(.6,-1.7){\footnotesize QD for GWLs $\ket{\psi_1}$}%
           }%
}%
\end{picture}
\caption{(color online) Plot of the EoF (line brown solid) and QD, in function of mixing
parameter $p$, for GWLs with a discrete pure state $\ket{\psi_{\max}}$, $\ket{\psi_3}$ (line
blue dashed), $\ket{\psi_2}$ (line magenta dotted) and $\ket{\psi_4}$ (line red dot-dashed)
those concurrences are equal to $C_{\max}=1$, $C_3=\tfrac{3}{4}$, $C_2=\tfrac{1}{2}$ and
$C_1=\tfrac{1}{4}$, respectively.} \label{fig:discretestates}
\end{figure}

\section{Applications}
In this section we show the EoF and QD for GWLs built with discrete and continue pure
states. For the discrete case we present four pure states with different entanglement,
and both measures of correlations are compared. In the other case, two mode $f-$deformed
coherent states are considered, and again the EoF and QD are compared.
\subsection{Discrete states}%
To illustrate the monotone behavior of the QD for the GWLs with respect to the
concurrence of the pure state with which are built, it is enough consider the following
states (in the representation of the matrix $\mathbb{W}_{\psi}$, given in the
Eqs.~\ref{eq:7})
\begin{equation}\label{eq:31}%
\begin{split}
&\mathbb{W}_{\Psi_{+}}=\tfrac{1}{\sqrt{2}}\begin{bmatrix} 1 & 0\\ 0 & 1\end{bmatrix},%
\hspace{.7cm}%
\mathbb{W}_{\psi_3}=\tfrac{1}{2\sqrt{10}}\begin{bmatrix} 3 & \hfill~\sqrt{6}\\ 2\sqrt{6} & -1\end{bmatrix},%
\\%
&\mathbb{W}_{\psi_2}=\tfrac{1}{6}\begin{bmatrix} -3 & -3\sqrt{2}\\ 2\sqrt{2} & 1\end{bmatrix},%
\quad
\mathbb{W}_{\psi_1}=\tfrac{1}{8}\begin{bmatrix} \sqrt{7} & \sqrt{5}\\ 3\sqrt{5} & \sqrt{7}\end{bmatrix}.%
\end{split}
\end{equation}
These matrices are representative of the equivalence classes corresponding to the concurrence
$1$, $\tfrac{3}{4}$, $\tfrac{1}{2}$ and $\tfrac{1}{4}$, respectively. The first matrix
corresponds to the Bell state $\ket{\Psi_{+}}$ (maximally entanglement), while the states
$\ket{\psi_1}$, $\ket{\psi_2}$ and $\ket{\psi_3}$ have less entanglement. In the
Fig.~\ref{fig:discretestates}, the EoF \eqref{eq:33} and QD \eqref{eq:30} have been sketched
for the four states given in \eqref{eq:31}. It can be observed that the QD is a monotonous
function that grows with the increase of the concurrence of the pure state associated to the
GWLs, but the QD not is a monotonous function of your own EoF, as it happens for pure states.
On the other hand, exists a region in which the QD is bigger than the EoF.

\subsection{$f-$Deformed Coherent States}%
The coherent states for the electromagnetic field, introduced by Glauber in
1963~\cite{Glaubera,Glauberb,Glauberc}, have played an important role, not only in quantum
optics, but in many fields of the physics. In terms of their evolution, they remain localized
about the corresponding classical trajectory when acted on by harmonic interactions and do not
change their functional form with the time~\cite{Gazeaua}. Glauber showed that these states
can be obtained from any one of these three mathematical definitions:
\begin{enumerate}
\item As the eigenstate of the annihilation bosonic operator $\hat{a}$, i.e.,
$\hat{a}\ket{\alpha}=\alpha\ket{\alpha}$, being $\alpha$ a complex number.
\item As those that can be obtained by the application of the displacement operator
$\hat{D}(\alpha)=\exp^{(\alpha\hat{a}^{\dag}-\alpha^{*}\hat{a})}$ on the vacuum state of
the harmonic oscillator, i.e., $\ket{\alpha}=\hat{D}(\alpha)\ket{0}$.
\item As the quantum states with a minimum uncertainty relationship
$(\triangle\hat{Q})^{2}_{\alpha}(\triangle\hat{P})^{2}_\alpha=1/4$, with $\hat{Q}$ and
$\hat{P}$ the position and momentum operators, respectively.
\end{enumerate}
Each of these definitions lead to equivalent coherent states for the harmonic
oscillators. On the other hand, two different coherent states are not orthogonal but they
form an complete basis, with which one can decompose any state~\cite{Gazeaub}.
\\
\par\indent%
The $f-$deformed oscillators are nonlinear oscillators with a specific kind of nonlinearity
for which the frequency depends on the oscillator energy, these single mode nonlinear states
are essentially based on the deformation of bosonic annihilation and creation operators,
according to the relations
\begin{subequations}\label{eq:58}
\begin{align}
\label{eq:58a}%
\hat{A}&=\hat{a}f(\hat{n})=f(\hat{n}+\hat{\openone})\hat{a},%
\\
\label{eq:58b}%
\hat{A}^{\dag}&=f(\hat{n})\hat{a}^{\dag}=\hat{a}^{\dag}f(\hat{n}+\hat{\openone}),
\end{align}
\end{subequations}
where $\hat{n}=\hat{a}^{\dag}\hat{a}$ is the bosonic number operator, with actions on the
Fox space as
\begin{subequations}\label{eq:59}
\begin{align}
\label{eq:59a}%
\hat{A}\ket{n}&=\sqrt{n}f(n)\ket{n-1},%
\\
\label{eq:59b}%
\hat{A}^{\dag}\ket{n}&=\sqrt{n+1}f(n+1)\ket{n+1}.
\end{align}
\end{subequations}
These deformed boson creator $\hat{A}$ and annihilator $\hat{A}^{\dag}$ differ from the
usual harmonic operators $\hat{a}$ and annihilation $\hat{a}^{\dag}$ by a deformation
function $f(\hat{n})$. This deformation function is real and non-negative, and it is
convenient to assume that is a continuous function with $f(0)=1$. The commutation
relations between the deformed operators are given by
\begin{equation}\label{eq:60}
[\hat{n},\hat{A}]=-\hat{A},\hspace{.2cm}%
[\hat{n},\hat{A}^{\dag}]=-\hat{A}^{\dag}, \hspace{.2cm}
[\hat{A},\hat{A}^{\dag}]=\hat{\openone}+\phi(\hat{n}).
\end{equation}
Where $\phi(\hat{n})=(\hat{n}+\hat{\openone}) f^{2}(\hat{n}+\hat{\openone}) - \hat{n}
f^{2}(\hat{n})-\hat{\openone}$. The deformation becomes fixed when one choose the
explicit form of the function $f(\hat{n})$, and the harmonic case is recovered when
$f(\hat{n})= \hat{\openone}$. The Hamiltonian of these deformed oscillators can be
written in terms of annihilation and creation deformed operators $\hat{A}$ and
$\hat{A}^{\dag}$ as
\begin{equation}\label{eq:61}
\widehat{H}=\frac{\hbar\Omega}{2}[\hat{A}\hat{A}^{\dag}+\hat{A}^{\dag}\hat{A}]\ ,
\end{equation}
and the spectrum of $\widehat{H}$ is given by
\begin{equation}\label{eq:62}
E_{n}=\frac{\hbar\Omega}{2}\left[(n+1)f^{2}(n+1)+nf^{2}(n)\right].
\end{equation}
\par\indent%
Man'ko and collaborators~\cite{Mankoa} introduced the nonlinear coherent states
$\ket{\alpha}_{A}$ of an $f-$oscillator algebra, as the eigenstates of the annihilation
operator $\hat{A}$, such as $\hat{A}\ket{\alpha}_{A}=\alpha\ket{\alpha}_{A}$. The explicit
form of normalized states $|\alpha\rangle_{A}$, in the number state representation, is given
by
\begin{subequations}\label{eq:63}
\begin{gather}
\label{eq:63a}%
\ket{\alpha}_{A}=N_{A}\sum_{n=0}^{\infty}\frac{\alpha^{n}}{\sqrt{n!}f(n)!}\ket{n},%
\intertext{with}%
\label{eq:63b}%
N_{A}=\left(\sum_{n=0}^{\infty}\frac{|\alpha|^{2n}}{n!(f(n)!)^{2}}\right)^{-1/2},
\end{gather}
\end{subequations}
where $f(n)!\definido f(0)f(1)f(2)\cdots f(n)$. When using the displacement operator
method to generate the coherent states for deformed algebra, one faces the problem that
the commutation between the deformed operators $\hat{A}$ and  $\hat{A}^{\dag}$ is not a
number, as a consequence, the displacement operator obtained by the replacement of the
usual operators, by their deformed counter parts can not be written in a product
form~\cite{Aniello}.
\\
\par\indent%
R\'{e}camier and collaborators~\cite{Recamiera} proposed an approach to generate the $f-$deformed
coherent state $\ket{\alpha}_{D}$ by the application of a deformed operator $\hat{D}(\alpha,f)$
acting upon the vacuum state, such that $\ket{\alpha}_{D}=\hat{D}(\alpha,f)\ket{0}$. The
R\'{e}camier deformed displacement operator can be written as
\begin{equation}\label{eq:47}%
\hat{D}(\alpha,f)=\exp^{\alpha\hat{A}^\dag}\exp^{\alpha^{*}\hat{A}}
\exp^{\frac{|\alpha|^{2}}{2}\,\left(1+\phi(\hat{n})\right)}.
\end{equation}
This displacement operator is nearly unitary and displaces $\hat{A}$ and $\hat{A}^{\dag}$
whenever
\begin{equation}\label{eq:64}
\begin{split}
&\hspace{2.5cm}\tfrac{1}{2}|\alpha|^2\phi(n)\ll1,%
\\
&\tfrac{1}{2}|\alpha|^2\left[(n+1)f^2(n+1)-nf^2(n)-1\right]\ll1.
\end{split}
\end{equation}
This restriction requieres that
the values of $n$ do not belong to the range $[0,\infty)$, but rather $n$ has a maximum
value $n_{\max}$ determined by this restriction on $|\alpha|^{2}$. The normalized
$f-$deformed coherent states obtained by application of the approximately displacement
operator upon the vacuum state are
\begin{subequations}\label{eq:27}%
\begin{gather}
\label{eq:27a}%
\ket{\alpha}_{D}=N_{D}\sum_{n=0}^{n_{\max}}\frac{\alpha^{n}\,f(n)!}{\sqrt{n!}}\ket{n},
\intertext{with}%
\label{eq:27b}%
N_{D}=\left(\sum_{n=0}^{n_{\max}}\frac{|\alpha|^{2n}\,(f(n)!)^{2}}{n!}\right)^{-1/2}.
\end{gather}
\end{subequations}
The deformed states $\ket{\alpha}_{A}$ and $\ket{\alpha}_{D}$ are not equivalents, in
general, they present a similar quantum evolution and however they show a different
statistical behavior~\cite{Recamierb}.
\\
\par\indent%
The QD and the EoF of the BWLs in the orthogonal basis $\ket{0}$ and $\ket{1}$ are well
known~\cite{LuoPRA77.042303.2008,Ali}. In this work, we are interested in studying the
behaviour of QD of the GWLs if we have bipartite entangled $f-$deformed coherent states.
We use the $|\alpha\rangle$, $|\alpha\rangle_{A}$ and $|\alpha\rangle_{D}$ coherent
states and $f-$deformed coherent states to encode qubits when they are superposed with
$\ket{-\alpha}$, $|-\alpha\rangle_{A}$ and $|-\alpha\rangle_{D}$, respectively. The
states $\ket{\alpha}_{X}$ and $\ket{-\alpha}_{X}$ correspond to non-orthogonal coherent
states (when $X=C$) and $f-$deformed coherent states (when $X=A$ or $X=D$) with opposite
phases. We choose an orthogonal basis by considering even and odd superpositions of
$|\alpha\rangle_{X}$ and $|\!-\alpha\rangle_{X}$, such that~\cite{Mann,Mishra}
\begin{subequations}\label{eq:46}%
\begin{gather}
\label{eq:46a}%
\ket{\pm}_X=\mathcal{N}^X_{\pm}\Big(\,\ket{\alpha}_{X}\pm\ket{-\alpha}_{X}\,\Big),%
\\
\label{eq:46b}%
\mathcal{N}^X_{\pm}=\big[2\pm2\,{}_X\!\braket{\alpha}{-\alpha}_X\big]^{-1/2},%
\intertext{and}
\label{eq:46c}%
{}_X\!\braket{\alpha}{-\alpha}_X=
\frac{\sum\limits_{n=0}^{n_{\max}}\frac{(-1)^n|\alpha|^{2n}}{n!}\left(f(n)!\right)^{\pm2}}
     {\sum\limits_{k=0}^{n_{\max}}\frac{|\alpha|^{2k}}{k!}\left(f(k)!\right)^{\pm2}},
\end{gather}
\end{subequations}
where $n_{\max}$ is taken according to the convergence of the expression \eqref{eq:64}
for $X=D$. The $+$ and $-$ signs in the equation \eqref{eq:46c} are taken when $X=D$ or
$X=A$, respectively. When $X=C$ then the deformation function is igual $1$ ($f(n)=1$),
with which $\ket{\alpha}_C$ is a coherent states $\ket{\alpha}$. These quantum
superpositions $\ket{\pm}_X$ can be considered as an realization of a $f-$deformed
Schr\"{o}dinger cat \cite{Recamierc}. In this paper, we are interested in to study the
Quasi-Bell entangled $f-$deformed coherent states even and odd and they are, respectively
\begin{subequations}\label{eq:2}
\begin{align}
\label{eq:2a}%
\ket{\Psi_+}_X&=n_X\left(\ket{\alpha,\alpha}_{X}+\ket{-\alpha,-\alpha}_{X}\right),%
\\
\label{eq:2b}%
\ket{\Psi_-}_X&=n_X\left( |\alpha,-\alpha\rangle_{X} + |-\alpha,\alpha\rangle_{X}\right),
\\
\label{eq:2c}%
n_X&=\left[2(1+|{}_X\!\braket{\alpha}{-\alpha}_{X}|^2)\right]^{-1/2}.%
\end{align}
\end{subequations}
If we express $\ket{\Psi_\pm}_X$ in the orthogonal basis of states $\ket{\pm}_{X}$ given
in \eqref{eq:46}, we can write
\begin{equation}\label{eq:3}
|\Psi_\pm\rangle_X=\frac{n_X}{2}\left[\frac{|+,+\rangle_X}{(\mathcal{N}^X_{+})^{2}}\pm\frac{|-,-\rangle_X}{(\mathcal{N}^X_{-})^{2}}\right]%
\end{equation}
These states are non-maximally entangled and are mutually non-orthogonal. When $f(n)=1$
and in the limit of large mean photon number $|\alpha|^{2}$ these states form a complete
orthogonal basis just like standard Bell states $|{\Psi}_{+}\rangle$ and
$|{\Psi}_{-}\rangle$. For the complete study of entanglement and quantum discord of GWLs
the matrix $\widehat{\mathbb{W}}_{\Psi_{\pm}^X}$ of the pure state \eqref{eq:3} is
requerid, which is diagonal and contains in the constants \eqref{eq:46b} and
\eqref{eq:2c} all the effects of the deformations. So that,
\begin{equation}\label{eq:13}
\widehat{\mathbb{W}}_{\Psi_{\pm}^X}=\frac{n_X}{2}%
\begin{bmatrix}
(\mathcal{N}^X_{+})^{-2} & 0\\
0                        & \pm(\mathcal{N}^X_{-})^{-2}
\end{bmatrix}.
\end{equation}
\begin{figure*}[t]
\begin{picture}(0,4.5)(3.5,0)
\put(-5,0){
\put(0,-.3){\includegraphics[height=5cm,width=7cm]{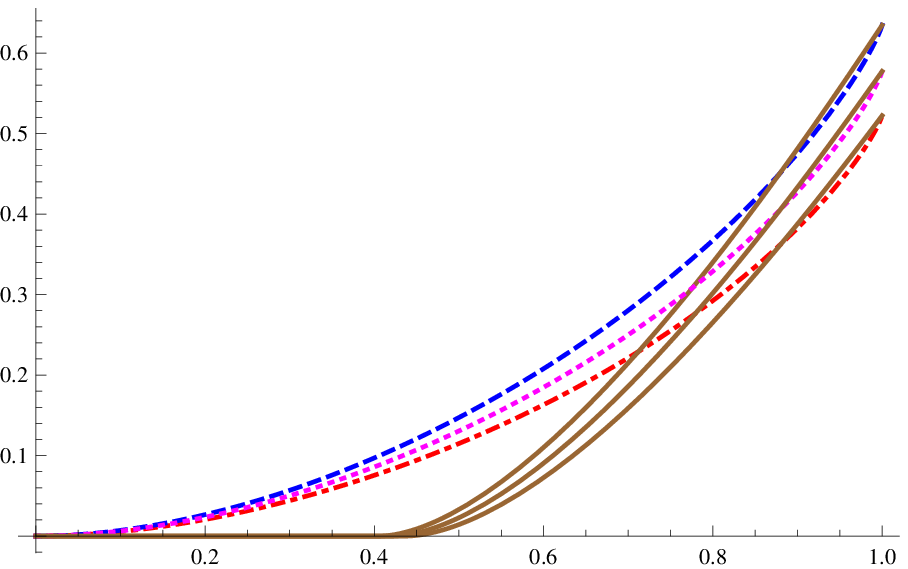}}%
\put(7.1,-.1){\footnotesize$p$}\put(.1,4.6){\footnotesize$EoF$, $\delta_{AB}$}%
\put(.7,3.4){%
 \put(-.2,.6){\parbox{4.5cm}{\footnotesize\centering%
           Fig.~\ref{fig:Poschl-Teller}(a)\\
           \underline{P\"{o}schl-Teller}\\
           $|\alpha|=.65$, $N=10$ and $n_{\max}=9$}}
 \psline[linecolor=brown](0,0)(.5,0)%
 \put(.6,-.1){\footnotesize EoF for $\ket{\alpha}_A$, $\ket{\alpha}_D$ and $\ket{\alpha}$.}%
 \multiput(0,-.4)(.13,0){4}{\psline[linecolor=magenta](0,0)(.07,0)}%
 \put(.6,-.5){\footnotesize QD for $\ket{\alpha}$}%
 \multiput(0,-.8)(.15,0){4}{\psline[linecolor=blue](0,0)(.1,0)}%
 \put(.6,-.9){\footnotesize QD for $\ket{\alpha}_A$}%
 \put(0,-1.2){\psline[linecolor=red](0,0)(.05,0)\psline[linecolor=red](.13,0)(.23,0)\psline[linecolor=red](.31,0)(.36,0)\psline[linecolor=red](.44,0)(.54,0)}%
 \put(.6,-1.3){\footnotesize QD for $\ket{\alpha}_D$}%
}%
}%
\put(4.3,0){
\put(0,-.3){\includegraphics[height=5cm,width=7cm]{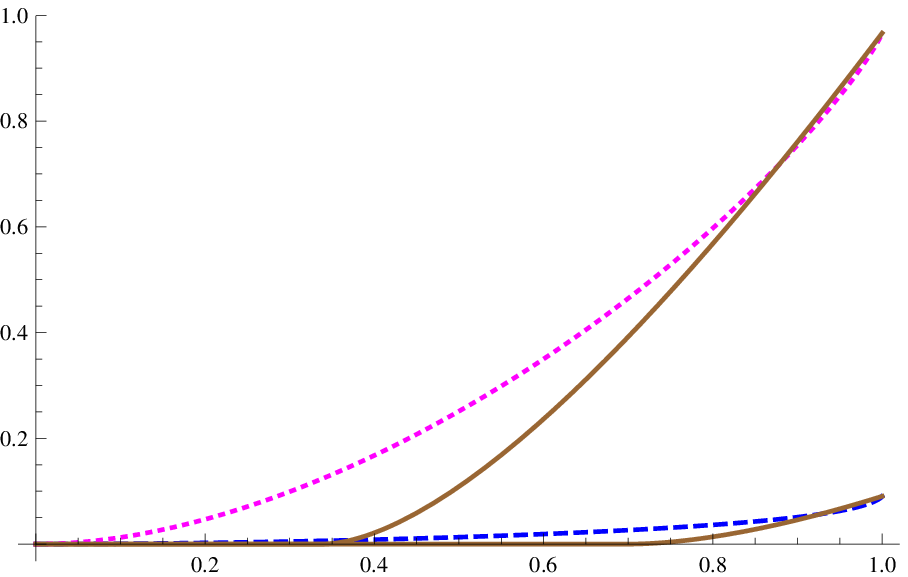}}%
\put(7.1,-.1){\footnotesize$p$}\put(.1,4.8){\footnotesize$EoF$, $\delta_{AB}$}%
\put(.7,3.4){%
 \put(-.2,.6){\parbox{4.5cm}{\footnotesize\centering%
           Fig.~\ref{fig:Poschl-Teller}(b)\\
           \underline{P\"{o}schl-Teller}\\
           $|\alpha|=3$, $N=10$ and $n_{\max}=9$}}
 \psline[linecolor=brown](0,0)(.5,0)%
 \put(.6,-.1){\footnotesize EoF for $\ket{\alpha}_A$ and $\ket{\alpha}$.}%
 \multiput(0,-.4)(.13,0){4}{\psline[linecolor=magenta](0,0)(.07,0)}%
 \put(.6,-.5){\footnotesize QD for $\ket{\alpha}$}%
 \multiput(0,-.8)(.15,0){4}{\psline[linecolor=blue](0,0)(.1,0)}%
 \put(.6,-.9){\footnotesize QD for $\ket{\alpha}_A$}%
 }%
}%
\end{picture}
\caption{(color online) Plot of the EoF (solid brown line) and QD, in function of mixing
parameter $p$, for the coherent state (dotted magenta line), $f-$deformed A like state (line
blue dashed) and $f-$deformed D like state (dot-dashed red line) linked to the
P\"{o}schl-Teller potential, when $N=10$ and $n_{\max}=9$. In figure (a) $|\alpha|=.65$
(small) and in the figure (b) $|\alpha|=3$ (large).} \label{fig:Poschl-Teller}
\end{figure*}
The $\widehat{\mathbb{W}}$ matrix for the $f-$deformed coherent states $\ket{\alpha}_A$ and
$\ket{\alpha}_D$ is diagonal in both cases, such that the GWLs are $X-$states. The QD of the
$X-$states has been studied numericallity~\cite{Girolami} and
analitically~\cite{LuoPRA77.042303.2008,Ali,Maldonado,Yurischev}. The concurrence of these
states is given for,
\begin{equation}\label{eq:1}%
C[\Psi_{\pm}^X]=2|\det\widehat{\mathbb{W}}_{\Psi_{\pm}^X}|=\frac{1}{2}\left(\frac{n_X}{N_{+}^X
N_{-}^X}\right)^2
\end{equation}
The QD of the GWLs corresponding to the pure states $\ket{\Psi_{+}^X}$ and
$\ket{\Psi_{-}^X}$ are the same. As consequence, in this word only we take
$\ket{\Psi_{+}^X}$ for study EoF and QD of GWLs corresponding to several $f-$deformed
functions.
\\%

\par\indent%
In order to illustrate the relevance of our results, we considered some deformation functions
that are important in quantum computing and in quantum information processing. Ours first
example, we considered the center-of-mass motion of trapped ion. In such model, both the
``center-of-mass motional states'' and the electronic-states can be simultaneously coupled and
manipulated by light fields~\cite{Yazdanpanah,Harounia}. The trapped ion systems are useful to
study the quantum optical and quantum dynamical properties of quantum systems that are
approximately isolated from the environment, and the strong Coulomb forces between the ions
can be used to realize logical gate operations by coupling different qubits. For this reason,
trapped atomic ions are one of the leading candidate systems to construct a robust quantum
computer \cite{Harty}. Recently, experimental entanglement between remote ions in different
ion traps modules has been reported~\cite{Hucul}. The trapped ion in a modified
P\"{o}schl-Teller potential can be considered as a $f-$deformed oscillator with a specific
kind of the $f-$deformed Heisenberg-Weyl algebra \cite{Harounia}, in which the corresponding
deformation function has the form
\begin{equation}\label{eq:65}
f(n)\to f_N(n)=\left[\frac{\sqrt{N^{2}+1}-n}{N}\right]^{1/2}
\end{equation}
where $N$ is a dimensionless positive parameter that is associated with the depth of the trap.
The values of $N$ and $n$ are not independent and they satisfy the relation $n<\sqrt{N^2+1}$.
In the limiting case where $N\longrightarrow\infty$, we obtain $f_N(n)=1$, therefore the
energy levels are related by the deep of the trap, and finite rang trap has a finite
dimensional Hilbert space, which is important because it seems posible realize experiments in
order to study the Hilbert space size effects on these systems. In
Fig.~\ref{fig:Poschl-Teller}, the EoF and QD for this trapped ion in a modified
P\"{o}schl-Teller potential are shown.
\\%

\par
\begin{figure*}[t]
\begin{picture}(0,4.5)(3.5,0)
\put(-5,0){
\put(0,-.3){\includegraphics[height=5cm,width=7cm]{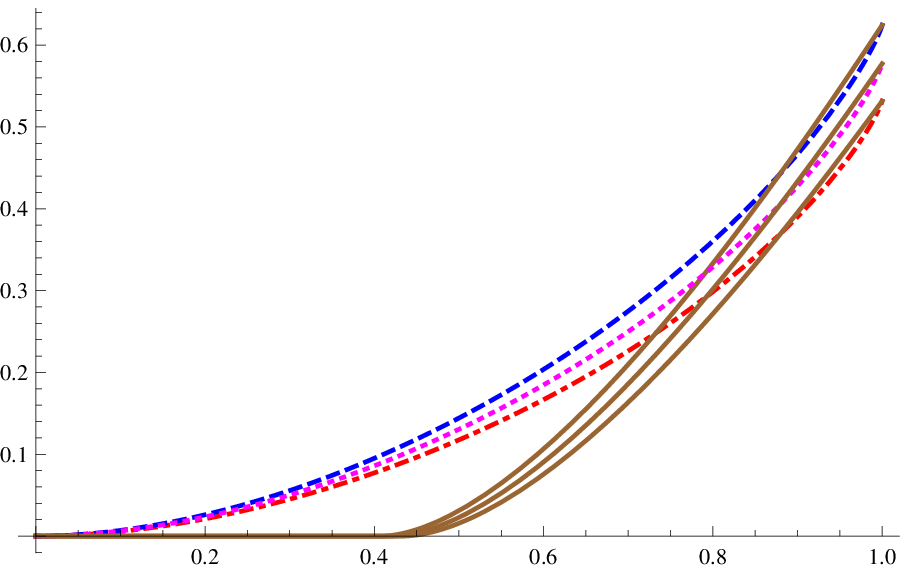}}
\put(7.1,-.1){\footnotesize$p$}\put(.1,4.6){\footnotesize$EoF$, $\delta_{AB}$}%
\put(.7,3.4){%
 \put(-.2,.6){\parbox{4.5cm}{\footnotesize\centering%
           Fig.~\ref{fig:Exciton}(a)\\
           \underline{Exciton}\\
           $|\alpha|=.65$, $\kappa=.3$ and $n_{\max}=5$}}
 \psline[linecolor=brown](0,0)(.5,0)%
 \put(.6,-.1){\footnotesize EoF for $\ket{\alpha}_A$, $\ket{\alpha}_D$ and $\ket{\alpha}$.}%
 \multiput(0,-.4)(.13,0){4}{\psline[linecolor=magenta](0,0)(.07,0)}%
 \put(.6,-.5){\footnotesize QD for $\ket{\alpha}$}%
 \multiput(0,-.8)(.15,0){4}{\psline[linecolor=blue](0,0)(.1,0)}%
 \put(.6,-.9){\footnotesize QD for $\ket{\alpha}_A$}%
 \put(0,-1.2){\psline[linecolor=red](0,0)(.05,0)\psline[linecolor=red](.13,0)(.23,0)\psline[linecolor=red](.31,0)(.36,0)\psline[linecolor=red](.44,0)(.54,0)}%
 \put(.6,-1.3){\footnotesize QD for $\ket{\alpha}_D$}%
 }%
           }%
\put(4.3,0){
\put(0,-.3){\includegraphics[height=5cm,width=7cm]{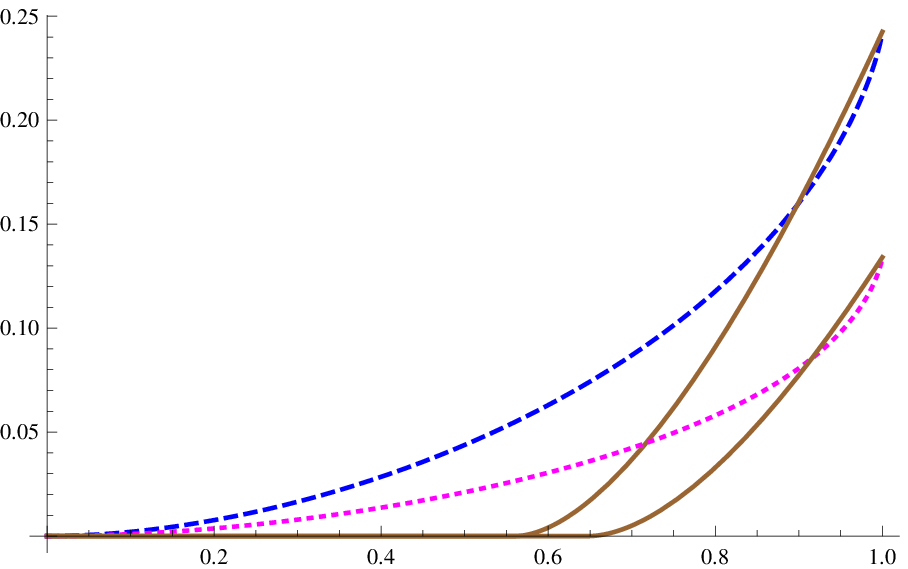}}%
\put(7.1,-.1){\footnotesize$p$}\put(.1,4.8){\footnotesize$EoF$, $\delta_{AB}$}%
\put(.7,3.4){%
 \put(-.2,.6){\parbox{4.5cm}{\footnotesize\centering%
           Fig.~\ref{fig:Exciton}(b)\\
           \underline{Exciton}\\
           $|\alpha|=6$, $\kappa=.3$ and $n_{\max}=5$}}
 \psline[linecolor=brown](0,0)(.5,0)%
 \put(.6,-.1){\footnotesize EoF for $\ket{\alpha}_A$ and $\ket{\alpha}$.}%
 \multiput(0,-.4)(.13,0){4}{\psline[linecolor=magenta](0,0)(.07,0)}%
 \put(.6,-.5){\footnotesize QD for $\ket{\alpha}$}%
 \multiput(0,-.8)(.15,0){4}{\psline[linecolor=blue](0,0)(.1,0)}%
 \put(.6,-.9){\footnotesize QD for $\ket{\alpha}_A$}%
}%
}%
\end{picture}
\caption{(color online) Plot of the EoF (solid brown line) and QD, in function of the mixing
parameter $p$, for the coherent state (dotted magenta line), A like $f-$deformed state (dashed
blue line) and D like $f-$deformed state (dot-dashed red line) linked to the exciton potential
in coupled quantum dots, when $\kappa=.3$. In figure (a) $|\alpha|=.6$ (small) and
$n_{\max}=3$, in figure (b) $|\alpha|=6$ (large) and $n_{\max}=8$.} \label{fig:Exciton}
\end{figure*}

\indent%
Due to the modern semiconductor microfabrication technology, the quantum dots are other of the
promising candidates for a solid-state quantum computer. These solid-state quantum systems are
especially attractive because of their good scalability and stability. Exciton in coupled
quantum dots are being used in the preparation of entangled states in solid-states systems
\cite{Quiroga}, and entangled of exciton states in a single quantum dot, or in a quantum dot
molecule, have been experimentally demonstrated \cite{Quiroga,Bayer,Stevenson,Jayakumar}.
Harouni and collaborators \cite{Harounib} proposed an $f-$deformed oscillator approach, to
study the confinement conditions of an exciton with definite angular momentum in a wide
quantum dot interacting with two lasers beans. Under this approach the deformation function
takes the form
\begin{equation}\label{eq:66}
f(n)\to f_\kappa(n)=\exp^{-\kappa^{2}}
\frac{L_{n}^{1}(\kappa^{2})}{(n+1)L_{n}^{0}(\kappa^{2})} \ ,
\end{equation}
where $\kappa$ is similar to the Lamb-Dicke parameter in trapped ion systems, and is defined
as the ratio of the quantum dot is radius to the wavelength of the driving laser, and the
$L_{m}^{n}(x)$ are the associated Laguerre Polynomials. This deformation function is similar
to the one that appears in the center-of-mass motion of a trapped ion confined in a harmonic
trap \cite{Aniello}, but in this case $\kappa$ is the Lamb-Dicke parameter which depend on the
laser wavelength and the quantum fluctuation of the ion position in the ground vibrational
state. For this deformation function the parameters $\kappa$ and $n$ not are independent, they
satisfy inequality $\kappa^2(2n+1)<<1$ in the Lamb-Dicke regime; this inequality bound to
maximum value of $n$ for $\kappa$ given. In Fig.~\ref{fig:Exciton} the EoF and QD for the
exciton potential in coupled quantum dots are shown.
\\
\par\indent%
Finally, we consider the entanglement of diatomic molecules.  The entanglement may play a
crucial role in explaining the relations of electronic and vibrational degrees of freedom in
molecules \cite{McKemmish}, and some diatomic molecules are the best candidates for multiple
molecular quantum bits in diatomic molecular quantum computers \cite{Ishii}. Molecular Quantum
computers may be more advantageous than atomic ion traps because the internal degrees of
freedom utilized as quantum bit for the former are much larger than those for the latter.
\\

\par\indent%
In order to model the entanglement between diatomic molecules, we consider the Morse potential
deformation function introduced by R\'{e}camier and
collaborators~\cite{Recamiera,Recamierb,Recamierc}. The Morse potential is an interatomic
model for the potential energy of a diatomic model, it is a better approximation for the
vibrational structure of these molecules than the quantum harmonic oscillator. The deformation
function of Morse potential has the form
\begin{equation}\label{eq:67}
f(n)\to f_N(n)=\sqrt{1+\frac{1-n}{2N}} \ ,
\end{equation}
where $N$ is an integer number determined by the number of bound states.  The values of $N$ and $n$
are not independent of each other, satisfying the relationship $n<\sqrt{2N+1}$. This deformation
function reproduce the spectra of the Morse and P\"{o}schl-Teller Hamiltonian \cite{Recamierb}. In
Fig.~\ref{fig:Morse} the EoF and QD for Morse potential are shown.

\section{Results}
\begin{figure*}[t]
\begin{picture}(0,4.5)(3.5,0)
\put(-5,0){
\put(0,-.3){\includegraphics[height=5cm,width=7cm]{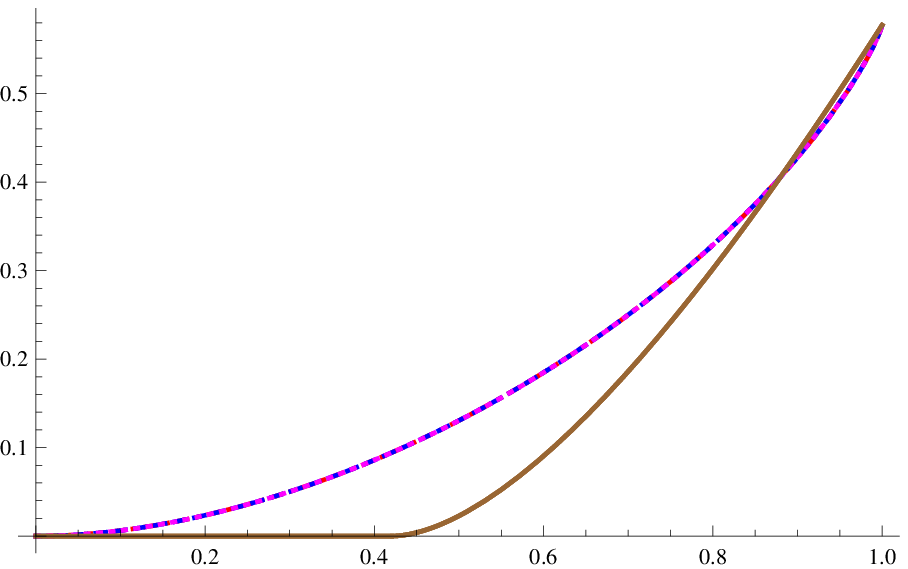}}%
\put(7.1,-.1){\footnotesize$p$}\put(.1,4.8){\footnotesize$EoF$, $\delta_{AB}$}%
\put(.7,3.4){%
 \put(-.2,.6){\parbox{4.5cm}{\footnotesize\centering%
           Fig.~\ref{fig:Morse}(a)\\
           \underline{Morse}\\
           $|\alpha|=.65$, $N=18$ and $n_{\max}=6$}}
 \psline[linecolor=brown](0,0)(.5,0)%
 \put(.6,-.1){\footnotesize EoF for $\ket{\alpha}_A$, $\ket{\alpha}_D$ and $\ket{\alpha}$.}%
 \multiput(0,-.4)(.13,0){4}{\psline[linecolor=magenta](0,0)(.07,0)}%
 \put(.6,-.5){\footnotesize QD for $\ket{\alpha}$}%
 \multiput(0,-.8)(.15,0){4}{\psline[linecolor=blue](0,0)(.1,0)}%
 \put(.6,-.9){\footnotesize QD for $\ket{\alpha}_A$}%
 \put(0,-1.2){\psline[linecolor=red](0,0)(.05,0)\psline[linecolor=red](.13,0)(.23,0)\psline[linecolor=red](.31,0)(.36,0)\psline[linecolor=red](.44,0)(.54,0)}%
 \put(.6,-1.3){\footnotesize QD for $\ket{\alpha}_D$}%
}%
}%
\put(4.3,0){
\put(0,-.3){\includegraphics[height=5cm,width=7cm]{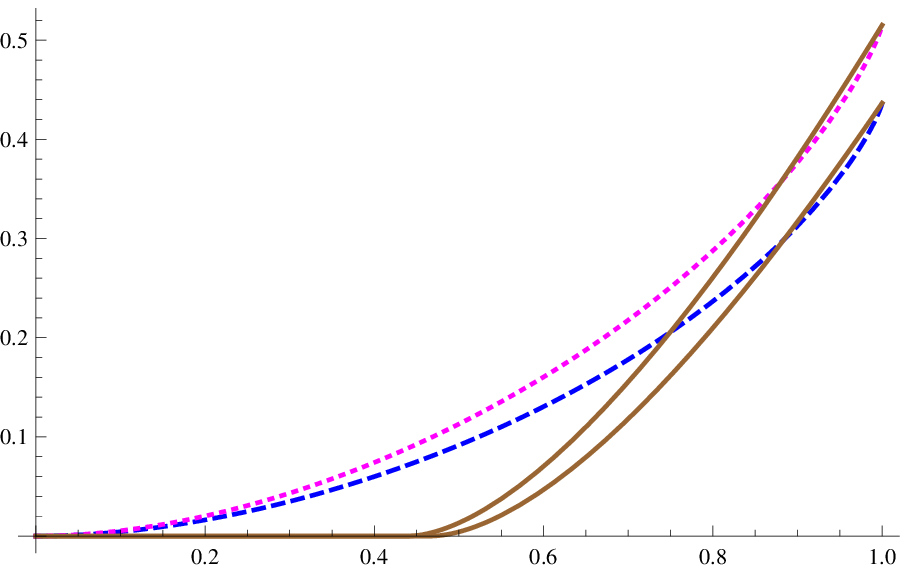}}%
\put(7.1,-.1){\footnotesize$p$}\put(.1,4.8){\footnotesize$EoF$, $\delta_{AB}$}%
\put(.7,3.4){%
 \put(-.2,.6){\parbox{4.5cm}{\footnotesize\centering%
           Fig.~\ref{fig:Morse}(b)\\
           \underline{Morse}\\
           $|\alpha|=4$, $N=18$ and $n_{\max}=6$}}
 \psline[linecolor=brown](0,0)(.5,0)%
 \put(.6,-.1){\footnotesize EoF for $\ket{\alpha}_A$ and $\ket{\alpha}$.}%
 \multiput(0,-.4)(.13,0){4}{\psline[linecolor=magenta](0,0)(.07,0)}%
 \put(.6,-.5){\footnotesize QD for $\ket{\alpha}$}%
 \multiput(0,-.8)(.15,0){4}{\psline[linecolor=blue](0,0)(.1,0)}%
 \put(.6,-.9){\footnotesize QD for $\ket{\alpha}_A$}%
 }%
}%
\end{picture}
\caption{(color online) Plot of the EoF (solid brown line) and QD, in function of the
mixing parameter $p$, for the coherent state (dotted magenta line), A like $f-$deformed
state (dashed blue line) and D like $f-$deformed state (dot-dashed red line) linked to
the Morse potential, when $N=18$ and $n_{\max}=10$. In figure (a) $|\alpha|=.65$ (small)
and in figure (b) $|\alpha|=4$ (large).} \label{fig:Morse}
\end{figure*}

In the previous section, we studied the behavior of the QD and EoF for continuous variable states
in their discrete formulations, taking as examples the coherent states and type $A$ and $D$
$f-$deformed coherent states, associated to deformations functions: the P\"{o}schl-Teller
(Fig.~\ref{fig:Poschl-Teller}); excitons coupled to a quantum dot (Fig.~\ref{fig:Exciton}); Morse
potential (Fig.~\ref{fig:Morse}). The QD and EoF are plotted for average numbers of photons
$|\alpha|$ smaller and greater than unity, observing that the EoF is greater than the QD for
certain values of $|\alpha|$. The graphs of the QD and EoF for states the type $D$ with $|\alpha|$
greater than unity are not shown, because they do not meet the inequality \eqref{eq:64} for the
values of $N$, $\kappa$ and $n_{\max}$ indicated in each figure. On the other hand, the difference
between our results and the numerical results of QD obtained using method in the
reference~\cite{Girolami}, for all the deformations functions considered, presented a percentage
relative error of $12$ significant digit. Moreover, for guarantee the convergence of the QD and EoF
are taken the value of $n_{\max}$ for which the QD (or the EoF) in $n_{\max}$ and $n_{\max}+1$  be
the same.
\\
\par\indent%
In this case of deformation function associated to P\"{o}schl-Teller the value of $N$ is fixed
to $10$, implies that the value of $n_{\max}$ must be less than $20$. However, the value taken
for $n_{\max}$ is $9$, because it guarantees the convergence of the QD and EoF; it means that,
for this value of $9$ the QD (or the EoF) is same to that obtained with the value of $10$.
Using the values of $|\alpha|=.65$, $N=10$ and $n_{\max}=9$ the restriction condition
\eqref{eq:64} takes the value of $0.4$. In other hand, the QD is greater that the EoF for the
states $\ket{\Psi_+}_A$ and $\ket{\Psi_+}_D$ when $|\alpha|>0.8785$ and $|\alpha|>0.8807$,
respectively. Furthermore, for values with $0<|\alpha|<1.3$ the QD of the state
$\ket{\Psi_+}_A$ is greater than the QD of the coherent state $\ket{\Psi_+}_C$, and the QD of
this state are greater than the QD of the state $\ket{\Psi_+}_D$, so that for this values of
$|\alpha|$ the QD satisfies the inequality (see Fig.~\ref{fig:Poschl-Teller}a)
\begin{equation}\label{eq:68}%
\delta_{AB}(\Psi_+^A,p)>\delta_{AB}(\Psi_+^C,p)>\delta_{AB}(\Psi_+^D,p).
\end{equation}
And for $|\alpha|>1.3$ we have (see Fig.~\ref{fig:Poschl-Teller}b)
\begin{equation}\label{eq:69}%
\delta_{AB}(\Psi_+^C,p)>\delta_{AB}(\Psi_+^A,p).
\end{equation}
\par\indent%
For the case of the deformation function associated to excitons coupled to a quantum dot
the value of $\kappa$ is fixed to $0.3$ to consider Lamb-Dicke regime, with which the
value of $n_{\max}$ must be less than $5.06$. Using the values of $|\alpha|=.65$,
$\kappa=.3$ and $n_{\max}=5$ the restriction condition \eqref{eq:64} takes the value of
$0.4$. On the other hand, the QD is greater that the EoF for the states $\ket{\Psi_+}_A$
and $\ket{\Psi_+}_D$ when $|\alpha|>0.8786$ and $|\alpha|>0.8804$, respectively. Also,
the inequality \eqref{eq:68} is hold for all the range of values of average number of
photons $|\alpha|$ (see Fig.~\ref{fig:Exciton})
\\
\par\indent%
Finally, for case of the deformation function associated Morse potential the value of $N$ is
fixed to a number maximum of the confined levels in one molecule. In this work we considered
$N=18$ and for $n_{\max}$, and must be lesser than $6.08$. With the values $|\alpha|=.65$,
$N=18$ and $n_{\max}=6$ the restriction condition \eqref{eq:64} takes the value of $0.07$.
Using this values we find that the QD is robust (no deformations to the coherent states) for
values $|\alpha|$ less than one (see Fig.~\ref{fig:Morse}a), nevertheless, there are
deformations for values of $|\alpha|$ greater than one (see Fig.~\ref{fig:Morse}b).

\section{Conclusions}
This work, permited to obtained the exact analytical solutions of QD and EoF for the Ws and
GWLs. The Ws and GWLs are different quantum states, since we can not obtain $\rho_W$ from
$\rho_{GWL}$, and viceversa, using a unitary transformation between them. The values of the
mixture parameter $p$, for which we can make comparisons of the correlations present in the Ws
and GWLs, are bounded by $-1/3\leq p<1/3$, finding that in this region both states are a
convex sum of product states. The QD and the EoF, for the GWLs are monotonic functions of the
concurrence of the pure state $\hat{\mathbb{P}}_{\psi}=\ket{\psi}\bra{\psi}$, and for pure
states with the same concurrence the  $QD=EoF$, for GWLs. On the other hand, the QD of the Ws
and GWLs are symmetric because their balance is zero.
\\
\par\indent%
To illustrated the relevance of our analytical results we calculated the QD and EoF for
$f-$deformed coherent states, in a discrete formulation of GWLs, applied to three different
deformed function: Post-teller, quantum dot excitons and Morse potentials. We find that the
difference between our results and the numerical results of QD, in all the considered cases,
presented a percentage relative error of $12$ significant figures. Regardless of the
deformation function used, Quasi-Bell states entangled $f-$deformed coherent states even and
odd have the same QD.
\\
\par\indent%
For the case of P\"{o}schl-Teller and the Morse potentials, there is a critical value for the
quantum number of photons $\alpha_c$, for which if $\alpha<\alpha_c$ the QD satisfies the
inequality $\delta_{AB}(\Psi_+^A,p)>\delta_{AB}(\Psi_+,p)>\delta_{AB}(\Psi_+^D,p)$, and for
$\alpha>\alpha_c$ the QD satisfies $\delta_{AB}(\Psi_+,p)>\delta_{AB}(\Psi_+^A,p)$. However, in the
case of the quantum dot potential, the QD satisfies the inequality
$\delta_{AB}(\Psi_+^A,p)>\delta_{AB}(\Psi_+,p)>\delta_{AB}(\Psi_+^D,p)$, for all $\alpha$ values.
The QD of the $f-$deformed coherent states $\ket{\alpha}_D$, obtained from the displacement
operator \eqref{eq:47} do not present QD for large values of $\alpha$ because the inequality
\eqref{eq:64} is not met.

\begin{acknowledgments}
We are especially thankful to Prof.~Manuel Rodriguez, Dr.~Jacinto Liendo and
Miguel Casanova for his fruitful discussions, comments and remarks on the final
version this article.
\end{acknowledgments}

\appendix
\section{Projective measurement onto pure state, Ws and GWLs}%
\label{sec:apenA}%
Let $\mathbb{U}=[U_{ij}]$ be unitary transformations, the base $\{\ket{\pi_m}\}$ is unitarily
equivalent to the computational base $\{\ket{i}\}$ if $\ket{\pi_m}=\sum_iU_{mi}\ket{i}$. The
projector, associated to these measurement are
\begin{equation}\label{eq:apenA1}%
\widehat{\Pi}_m=\ket{\pi_m}\bra{\pi_m}=\sum_{ij}U_{im}\overline{U}_{jm}\ket{i}\bra{j},
\end{equation}
where $\overline{U}_{jm}$ is the complex conjugate of $U_{jm}$. The projectors associated to
local projective measurement in the partition $A$ of bipartite systems are
\begin{equation}\label{eq:apenA2}%
\widehat{\Pi}_m^A=\widehat{\Pi}_m\otimes\hat{\openone}=%
\sum_{ijk}U_{im}\overline{U}_{jm}\ket{ik}\bra{jk},
\end{equation}
where the identity operator $\hat{\openone}$ has been replaced by the sum of projectors
$\sum_k\ket{k}\bra{k}$. In the other hand, any pure state \ket{\psi} belonging to
$\mathscr{H}\otimes\mathscr{H}$ can be written in term of the computational base as
\begin{equation}\label{eq:apenA3}%
\ket{\psi}=\sum_{ij}\psi_{ij}\ket{ij}%
\quad\textrm{with}\quad \sum_{ij}\psi_{ij}\overline{\psi}_{ij}=1.
\end{equation}
In order to simplify our results we define the matrix $\widehat{\mathbb{W}}_\psi$, whose
elements are $\psi_{ij}$, so the normalization condition can be written as
\begin{gather}
\sum_{ij}\psi_{ij}\overline{\psi}_{ij}=1\;\Longrightarrow\;
\sum_{ij}[\widehat{\mathbb{W}}_\psi]_{i\times j}[\overline{\widehat{\mathbb{W}}_\psi}]_{i\times j}=1%
\nonumber\\
\sum_{ij}[\widehat{\mathbb{W}}_\psi]_{i\times j}[\widehat{\mathbb{W}}_\psi^\dag]_{j\times i}=1
\;\Longrightarrow\;%
\sum_{i}[\widehat{\mathbb{W}}_\psi\widehat{\mathbb{W}}_\psi^\dag]_{i\times i}=1%
\nonumber\\
\label{eq:apenA4}%
\therefore\;
\tr{\widehat{\mathbb{W}}_\psi\widehat{\mathbb{W}}_\psi^\dag}=\tr{\widehat{\mathbb{W}}_\psi^\dag\widehat{\mathbb{W}}_\psi}=1
\end{gather}
The representation of pure states in terms of density matrix is given by the following
projector of rank one,
\begin{equation}\label{eq:apenA6}%
\ket{\psi}\bra{\psi}=\sum_{ijk\ell}\psi_{ij}\overline{\psi}_{k\ell}\ket{ij}\bra{k\ell},
\end{equation}
the reduced state are obtained taking partial trace over both partitions, so for partition $A$
we have that
\begin{gather}
\rho_A(\psi)=\tr[B]{\ket{\psi}\bra{\psi}}=
\sum_{ijk\ell}\psi_{ij}\overline{\psi}_{k\ell}\tr[B]{\ket{ij}\bra{k\ell}}%
\nonumber\\%
=\sum_{ijk\ell}\psi_{ij}\overline{\psi}_{k\ell}\delta_{j\ell}\ket{i}\bra{k}
=\sum_{ijk}\psi_{ij}\overline{\psi}_{kj}\ket{i}\bra{k}.
\nonumber\\%
=\sum_{ijk}[\widehat{\mathbb{W}}_\psi]_{i\times j}[\widehat{\mathbb{W}}_\psi^\dag]_{j\times k}\ket{i}\bra{k}%
=\sum_{ik}[\widehat{\mathbb{W}}_\psi\widehat{\mathbb{W}}_\psi^\dag]_{i\times k}\ket{i}\bra{k}.%
\nonumber\\%
\label{eq:apenA7}%
\therefore\quad\rho_A(\psi)=\widehat{\mathbb{W}}_\psi\widehat{\mathbb{W}}_\psi^\dag
\end{gather}
and for partition $B$ we have,
\begin{gather}
\rho_B(\psi)=\tr[A]{\ket{\psi}\bra{\psi}}=
\sum_{ijk\ell}\psi_{ij}\overline{\psi}_{k\ell}\tr[A]{\ket{ij}\bra{k\ell}}%
\nonumber\\%
=\sum_{ijk\ell}\psi_{ij}\overline{\psi}_{k\ell}\delta_{ik}\ket{j}\bra{\ell}
=\sum_{ij\ell}\psi_{ij}\overline{\psi}_{i\ell}\ket{j}\bra{\ell}.
\nonumber\\%
=\sum_{ij\ell}[\widehat{\mathbb{W}}_\psi]_{i\times j}[\widehat{\mathbb{W}}_\psi^\dag]_{\ell\times i}\ket{j}\bra{\ell}%
=\sum_{ij\ell}[\widehat{\mathbb{W}}_\psi^T]_{j\times i}[(\widehat{\mathbb{W}}_\psi^T)^\dag]_{i\times\ell}\ket{j}\bra{\ell}%
\nonumber\\%
=\sum_{j\ell}[(\widehat{\mathbb{W}}_\psi^T)(\widehat{\mathbb{W}}_\psi^T)^\dag]_{j\times \ell}\ket{j}\bra{\ell}.%
\nonumber\\%
\label{eq:apenA8}%
\quad\therefore\quad\rho_B(\psi)=(\widehat{\mathbb{W}}_\psi^T)(\widehat{\mathbb{W}}_\psi^T)^\dag
\end{gather}
This shows that partition $B$ can be accesses through the transpose operation. On the other
hand, the probability of obtaining a result after the local projective measurement
\eqref{eq:apenA2} when the system is initially in the pure state $\ket{\psi}$ is give by
\begin{equation}\label{eq:apenA5}%
\vmedio{\widehat{\Pi}_m^A}_\psi=\bra{\psi}\widehat{\Pi}_m^A\ket{\psi}=%
\sum_{ijk}\overline{\psi}_{ik}\psi_{jk}U_{im}\overline{U}_{jm},
\end{equation}
in this expression have been replace \eqref{eq:apenA2}. The last expression can be written in
matrix form as,
\begin{gather}
\vmedio{\widehat{\Pi}_m^A}_\psi=\sum_{ijk}\overline{\psi}_{ik}U_{im}\overline{U}_{jm}\psi_{jk}%
\nonumber\\
\sum_{ijk}
[\widehat{\mathbb{W}}_\psi^\dag]_{k\times i}[\widehat{\Pi}_m]_{i\times j}[\widehat{\mathbb{W}}_\psi]_{j\times k}%
=\tr{\widehat{\mathbb{W}}_\psi^\dag\widehat{\Pi}_m\widehat{\mathbb{W}}_\psi}
\nonumber\\
\label{eq:apenA9}%
\vmedio{\widehat{\Pi}_m^A}_\psi=
\tr{\widehat{\mathbb{W}}_\psi\widehat{\mathbb{W}}_\psi^\dag\widehat{\Pi}_m}%
\equiv\vmedio{\widehat{\Pi}_m}_{\rho_A(\psi)},
\end{gather}
where we have used the expressions \eqref{eq:apenA1} and \eqref{eq:apenA7}. If the measurement is
performed on partition $B$ then
\begin{equation}\label{eq:apenA10}%
\vmedio{\widehat{\Pi}_m^B}_\psi=\vmedio{\widehat{\Pi}_m}_{\rho_B(\psi)}=%
\tr{(\widehat{\mathbb{W}}_\psi^T)(\widehat{\mathbb{W}}_\psi^T)^\dag\widehat{\Pi}_m}.%
\end{equation}
We perform a projective measurement on the pure state on partition $A$. The state after of the
measure is obtains by L\"uders rule~\cite{luders}, so
\begin{gather}
\label{eq:apenA15}
\ket{\psi}\bra{\psi}\Big|_{\Pi_m^A}=%
\frac{\widehat{\Pi}_m^A\ket{\psi}\bra{\psi}(\widehat{\Pi}_m^A)^\dag}{\vmedio{\widehat{\Pi}_m^A}_\psi}%
\\%
=\tfrac{1}{\vmedio{\widehat{\Pi}_m^A}_\psi}\sum_{ijk}\sum_{rst}%
\psi_{jk}\overline{\psi}_{st}U_{sm}U_{im}\overline{U}_{jm}\overline{U}_{rm}\ket{ik}\bra{rt}
\nonumber\\%
=\tfrac{1}{\vmedio{\widehat{\Pi}_m^A}_\psi}\left[\sum_{ir}U_{im}\overline{U}_{rm}\ket{i}\bra{r}\right]%
\otimes\left[\sum_{jkst}\overline{\psi}_{st}U_{sm}\overline{U}_{jm}\psi_{jk}\ket{k}\bra{t}\right]%
\nonumber\\%
=\tfrac{1}{\vmedio{\widehat{\Pi}_m^A}_\psi}\widehat{\Pi}_m%
\otimes\left[\sum_{jkst}[\widehat{W}_\psi^\dag]_{t\times s}[\widehat{\Pi}_m]_{s\times j}[\widehat{W}_\psi]_{j\times k}\ket{k}\bra{t}\right]%
\nonumber\\%
=\widehat{\Pi}_m\otimes\left[\sum_{kt}\frac{[\widehat{W}_\psi^\dag\widehat{\Pi}_m\widehat{W}_\psi]_{t\times k}}{\vmedio{\widehat{\Pi}_m^A}_\psi}\,\ket{k}\bra{t}\right]%
\nonumber\\%
\label{eq:apenA11}\therefore\;%
\ket{\psi}\bra{\psi}\Big|_{\Pi_m^A}=%
\widehat{\Pi}_m\otimes\rho_{B|\Pi_m^A}
\end{gather}
where we have defined
\begin{equation}\label{eq:apenA12}%
\rho_{B|\Pi_m^A}=
\sum_{ij}\frac{\bra{i}\widehat{W}_\psi^\dag\widehat{\Pi}_m\widehat{W}_\psi\ket{j}}{\tr{\widehat{W}_\psi^\dag\widehat{\Pi}_m\widehat{W}_\psi}}\,\ket{j}\bra{i}%
\equiv\ket{\widehat{\psi}}\bra{\widehat{\psi}}.
\end{equation}
In the equation \eqref{eq:apenA12} it has been replace in the equation \eqref{eq:apenA9}. We
can show that \eqref{eq:apenA12} is pure state, since it is projector operator of rank one. A
straightforward calculator leads to
\begin{equation}\label{eq:apenA13}%
\tr{\rho_{B|\Pi_m^A}}=1\quad\textrm{and}\quad\rho_{B|\Pi_m^A}^2=\rho_{B|\Pi_m^A}.
\end{equation}
In the case of projective measurement in the partition $B$ the results are similar, except for
the transpose operation in the matrix $\widehat{\mathbb{W}}_\psi$.
\\
\par\indent%
Now we perform a local projective measure in the partition A to the Ws given in
\eqref{eq:0}. According to \eqref{eq:16a} we have that
\begin{gather}
\rho_{\textrm{W}|\Pi_{m}^{A}}=\frac{(\widehat{\Pi}_{m}^{A})\rho_{\textrm{W}}(\widehat{\Pi}_{m}^{A})^{\dag}}{p_{m}^{A}}%
\nonumber\\
\rho_{\textrm{W}|\Pi_{m}^{A}}%
=\tfrac{1}{p_{m}^{A}}\widehat{\Pi}_m^A\left[\tfrac{1-p}{4}\hat{\openone}_4+\tfrac{p}{2}\hat{\mathbb{F}}_4\right](\widehat{\Pi}_m^A)^\dag%
\nonumber\\
\rho_{\textrm{W}|\Pi_{m}^{A}}%
=\tfrac{1}{p_{m}^{A}}\left[\tfrac{1-p}{4}\widehat{\Pi}_m^A(\widehat{\Pi}_m^A)^\dag+\tfrac{p}{2}\widehat{\Pi}_m^A\hat{\mathbb{F}}_4(\widehat{\Pi}_m^A)^\dag\right]%
\nonumber\\
\rho_{\textrm{W}|\Pi_{m}^{A}}%
=\tfrac{1}{p_{m}^{A}}\left[\tfrac{1-p}{4}\widehat{\Pi}_m^A+\tfrac{p}{2}\sum_{ij}\widehat{\Pi}_m^A\ket{ij}\bra{ji}\widehat{\Pi}_m^A\right]%
\nonumber\\
\rho_{\textrm{W}|\Pi_{m}^{A}}%
=\tfrac{1}{p_{m}^{A}}\left[\tfrac{1-p}{4}\widehat{\Pi}_m^A+\tfrac{p}{2}\sum_{ij}(\widehat{\Pi}_m\ket{i}\bra{j}\widehat{\Pi}_m)\otimes(\ket{j}\bra{i})\right]%
\nonumber\\
\rho_{\textrm{W}|\Pi_{m}^{A}}%
=\tfrac{1}{p_{m}^{A}}\left[\tfrac{1-p}{4}\widehat{\Pi}_m^A+\tfrac{p}{2}\sum_{ij}U_{im}\bar{U}_{jm}(\ket{k_m}\bra{k_m})\otimes(\ket{j}\bra{i})\right]%
\nonumber\\
\rho_{\textrm{W}|\Pi_{m}^{A}}%
=\tfrac{1}{p_{m}^{A}}\left[\tfrac{1-p}{4}\widehat{\Pi}_m\otimes\hat{\openone}_2+\tfrac{p}{2}\widehat{\Pi}_m\otimes\widehat{\Pi}_m\right]%
\nonumber\\
\label{eq:apenA18}%
\rho_{\textrm{W}|\Pi_{m}^{A}}%
=\widehat{\Pi}_m\otimes\left[\frac{1-p}{4p_{m}^{A}}\hat{\openone}_2+\frac{p}{4p_m^A}\widehat{\Pi}_m\right]%
\end{gather}
Using \eqref{eq:38} in the equation \eqref{eq:apenA18} we obtain the equation
\eqref{eq:16b}.
\\
\par\indent%
Finally, we perform a local projective measurement on the GWLs \eqref{eq:4} in the partition
$A$. According to the equation \eqref{eq:14a} we have
\begin{gather}
\rho_{\textrm{GWL}|\Pi_{m}^{A}}=\frac{(\widehat{\Pi}_{m}^{A})\rho_{\textrm{GWL}}(\widehat{\Pi}_{m}^{A})^{\dag}}{p_{m}^{A}}%
\nonumber\\
\rho_{\textrm{GWL}|\Pi_{m}^{A}}%
=\tfrac{1}{p_{m}^{A}}\widehat{\Pi}_m^A\left[\tfrac{1-p}{4}\hat{\openone}_4+p\ket{\psi}\bra{\psi}\right](\widehat{\Pi}_m^A)^\dag%
\nonumber\\
\rho_{\textrm{GWL}|\Pi_{m}^{A}}%
=\tfrac{1}{p_{m}^{A}}\left[\tfrac{1-p}{4}\widehat{\Pi}_m^A(\widehat{\Pi}_m^A)^\dag+p\widehat{\Pi}_m^A\ket{\psi}\bra{\psi}(\widehat{\Pi}_m^A)^\dag\right]%
\nonumber\\
\rho_{\textrm{GWL}|\Pi_{m}^{A}}%
=\tfrac{1}{p_{m}^{A}}\left[\tfrac{1-p}{4}\widehat{\Pi}_m^A+p\vmedio{\widehat{\Pi}_m^A}_\psi\ket{\psi}\bra{\psi}\Big|_{\Pi_m^A}\right]%
\nonumber\\
\label{eq:apenA14}%
\rho_{\textrm{GWL}|\Pi_{m}^{A}}%
=\widehat{\Pi}_m\otimes\left[\frac{1-p}{4p_{m}^{A}}\hat{\openone}_2+\frac{p\vmedio{\widehat{\Pi}_m^A}_\psi}{p_m^A}\ket{\widehat{\psi}}\bra{\widehat{\psi}}\right]%
\end{gather}
Where we have be used Eqs.~\eqref{eq:apenA15}, \eqref{eq:apenA11} and \eqref{eq:apenA12}.
Defining the mixing parameter in the partition $B$ as
\begin{equation}\label{eq:apenA16}%
x_m(p)=\frac{p\vmedio{\widehat{\Pi}_m^A}_\psi}{p_m^A}=\frac{p\vmedio{\widehat{\Pi}_m^A}_\psi}{\tfrac{1-p}{2}+p\vmedio{\widehat{\Pi}_m^A}_\psi}.
\end{equation}
While the term that accompanies the identity matrix in \eqref{eq:14b} can be written as
\begin{equation}\label{eq:apenA17}%
\frac{1-p}{4p_{m}^{A}}=\frac{1-p}{4p\vmedio{\widehat{\Pi}_m^A}_\psi/x_m(p)}=\frac{1-x_m(p)}{2}.
\end{equation}
These showed the equation \eqref{eq:16}. This result is very important since the projective
measurement does no alter the structure of the GWLs, but modifies the mixing parameter $p$ by
$x_m(p)$.
\section{Calculation of condicional entropy for Werner-like states}%
\label{sec:apenB}%
For the optimization process, it is convenient to define \eqref{eq:23}, which is a positive an
monotonically increasing function of the mixing parameter $x_m(p)$, of partition $B$. So that
the conditional entropy \eqref{eq:20} is given by
\begin{equation}\label{eq:apenB1}%
S_{B|\{\Pi_m^A\}}(\psi,p)%
=\min_{\{\Pi_m^A\}}\sum_{m}F\left(x_m(p)\right).
\end{equation}
The minimum is obtained when there exist a set of value for the mixing parameter $x_m(p)$ such
that the function $F$ is minimal, subject to restriction \eqref{eq:18}. For the case $n=2$, it
is sufficient find the value $\underline{x}_0$ for which $F$ is minimal, while
$\underline{x}_1$ is obtained from \eqref{eq:18}. Deriving $F(z_m)$ with respect to $z_m$ and
after a simple calculation, we can obtain
\begin{equation}\label{eq:apenB2}%
dF(x_m)=-\frac{(1-p)\log_2\left(\tfrac{1+x_m}{2}\right)}{2(1-z_m)^2}\,dx_m.
\end{equation}
Using the values of $x_{m}$ given in \eqref{eq:16}, it is easy to show that
\begin{equation}\label{eq:apenB3}%
dF(x_m)=p\log_2\left[\frac{\tfrac{1-p}{2}+p\vmedio{\widehat{\Pi}^A_m}_\psi}{\tfrac{1-p}{4}+p\vmedio{\widehat{\pi}^A_m}_\Psi}\right]\,d\vmedio{\widehat{\Pi}^A_m}_\psi.%
\end{equation}
\\*[-.5cm]
\par\noindent%
It is clear from \eqref{eq:apenB3} that the process of minimizing the conditional entropy is
relegated to finding the values of $x_{m}$ that minimize the probability
$\vmedio{\widehat{\Pi}^A_m}_\psi$, which in turn minimize the function $F(x_m)$. This
probability presents oscillations around the uniform distribution, which allows us to evaluate
its minimum quickly. Considering the local projective measurement \eqref{eq:8} and after
straightforward calculation, we obtain the simplified result
\begin{equation}\label{eq:apenB4}%
\begin{split}
\vmedio{\widehat{\Pi}_m^A}_\psi=\tfrac{1}{2}&\Big[1+\vmedio{\sigma_z}_{\rho_A(\psi)}\cos(2\theta+m\pi)%
\\%
&+\vmedio{\exp^{i\phi\sigma_z}\sigma_x}_{\rho_A(\psi)}\sin(2\theta+m\pi)\Big]%
\end{split}
\end{equation}
where $\rho_A(\psi)$ is given by \eqref{eq:apenA7} and the explicit expressions for the
coefficients of the trigonometric functions are
\begin{subequations}\label{eq:apenB5}%
\begin{eqnarray}
\label{eq:apenB5a}%
&&\hspace{-1cm}\vmedio{\sigma_z}_{\rho_A(\psi)}=\sqrt{|\psi_{00}|^2+|\psi_{01}|^2-|\psi_{10}|^2-|\psi_{11}|^2}\,,%
\\
\label{eq:apenB5b}%
&&\hspace{-1cm}\vmedio{\exp^{i\phi\sigma_z}\sigma_x}_{\rho_A(\psi)}=2\Re\left[(\psi_{00}\overline{\psi}_{10}-\psi_{01}\overline{\psi}_{11})\exp^{-i\phi}\right].
\end{eqnarray}
\end{subequations}
Taking into account that $\Re\left[z\exp^{i\phi}\right]\le|z|$ and
\begin{displaymath}
2|\psi_{00}\overline{\psi}_{10}-\psi_{01}\overline{\psi}_{11}|=%
\sqrt{\vmedio{\sigma_x}_{\rho_A(\psi)}^2+\vmedio{\sigma_y}_{\rho_A(\psi)}^2},
\end{displaymath}
we have that the amplitude of the oscillations presented in \eqref{eq:apenB4} is given by
\begin{equation}\label{eq:apenB6}%
A=\tfrac{1}{2}\sqrt{\vmedio{\sigma_z}_{\rho_A(\psi)}^2+\vmedio{\sigma_x}_{\rho_A(\psi)}^2+\vmedio{\sigma_y}_{\rho_A(\psi)}^2},
\end{equation}
This result coincides with \eqref{eq:25}. So the minimum probability value is
\begin{equation}
\vmedio{\widehat{\Pi}_m^A}_\psi^{\min}=\tfrac{1}{2}-A.
\end{equation}
Sintetizing, the value that minimize the function $F(x_m(p))$, and therefore minimize the
conditional entropy \eqref{eq:apenB1}, given by \eqref{eq:24a}.

\begin{small}

\end{small}

\end{document}